\documentclass[a4paper,11pt]{article}
\usepackage{jcappub}
\usepackage{epsfig}
\usepackage{graphics}
\usepackage{color}
\definecolor{orange}{rgb}{1,0.5,0}

\title{Ultrahigh Energy Cosmic Ray Nuclei from Extragalactic Pulsars and the effect of their Galactic counterparts}
\author[a]{Ke Fang}
\author[b]{, Kumiko Kotera}
\author[a]{and Angela V. Olinto}
\affiliation[a]{Department of Astronomy \& Astrophysics, Kavli Institute for Cosmological Physics, The
  University of Chicago, Chicago, Illinois 60637, USA.}
\affiliation[b]{Institut d'Astrophysique de Paris, UMR 7095 - CNRS, Universit\'e Pierre $\&$ Marie Curie, 98 bis boulevard Arago, 75014, Paris, France}

\abstract{The acceleration of ultrahigh energy nuclei in fast spinning newborn pulsars can explain the observed spectrum of ultrahigh energy cosmic rays and the trend towards heavier nuclei for energies above $10^{19}\,$eV as  reported by the Auger Observatory. Pulsar acceleration implies a hard injection spectrum ($\sim E^{-1}$) due to pulsar spin down and a maximum energy $E_{\rm max} \sim Z \, 10^{19}$ eV due to the limit on the spin rate of neutron stars. We have previously shown that the escape through the young supernova remnant softens the spectrum,  decreases slightly the maximum energy, and generates secondary nuclei. Here we show that the distribution of pulsar birth periods and the effect of propagation in the interstellar and intergalactic media modifies the combined spectrum of all pulsars. By assuming a normal distribution of pulsar birth periods centered at 300 ms, we show that the contribution of extragalactic pulsar births to the ultrahigh energy cosmic ray spectrum naturally gives rise to a contribution  to very high energy cosmic rays (VHECRs, between $10^{16}$ and $10^{18}$ eV) by Galactic pulsar births. The required injected composition to fit the observed spectrum depends on the absolute energy scale, which is uncertain, differing between Auger Observatory and Telescope Array. The contribution of Galactic pulsar births can also bridge the gap between predictions for cosmic ray acceleration in supernova remnants and the observed spectrum just below the ankle, depending on the composition of the cosmic rays that escape the supernova remnant and the diffusion behavior of VHECRs in the Galaxy.}
\keywords{cosmic ray acceleration, neutron star, pulsar, supernova, ultrahigh energy cosmic rays, very high energy cosmic rays}

\begin{document}

\maketitle 

\section{Introduction}

The origin of cosmic rays continues to challenge our understanding after a century of observations. Space and balloon based observatories have made precise measurements of the spectrum and composition of cosmic rays up to $\lesssim 10^{15}$ eV per particle. Above these energies, observatories on the ground have studied extensive air showers up to $10^{20}$ eV.  The bulk of the cosmic ray flux is believed to be accelerated in Galactic supernova remnants (SNR) \cite{Baade1934,Bell78,Blandford78}. This long held notion fits well the observed spectrum up to $10^{16}$ eV \cite{Blasi:2011fi}. Above these energies a new component is needed to explain the spectrum and observed composition. This new component may be Galactic, as suggested in \cite{Hillas:2006ms,Ptuskin:2010zn}, or extragalactic as proposed in \cite{Berezinsky:2005cq,L05}. Constraints derived from the lack of strong anisotropies in cosmic ray arrival directions and the observed structure of the Galactic Magnetic field show that above $2 \times 10^{19}$ eV cosmic rays are extragalactic \cite{Giacinti:2011ww}. However, the transition from Galactic to extragalactic is expected to occur well below $10^{19}$ eV, with models spanning the very high energy (VHE) range between  $10^{16}$ eV and $10^{18}$ eV with ``dip'' models around $10^{17}$ eV  \cite{Berezinsky:2005cq,L05} and ``ankle" transition models around $10^{18}$ eV  (see, e.g., \cite{Allard05}).

The study of ultrahigh energy cosmic rays (UHECRs), from $10^{18}$ eV to $10^{20}$ eV, has progressed significantly with the advent of giant airshower observatories such as the 3,000 km$^2$ Pierre Auger Observatory in Mendoza, Argentina \cite{Abraham04} and the 700 km$^2$ Telescope Array (TA) in Utah, USA \cite{AbuZayyad:2012kk,Tokuno:2012mi}. 
The spectrum, sky distribution of arrival directions, and composition indicators are well measured over a large range of energies. Differences in reports from the two major observatories include a 20\% shift in absolute energy scale ($E_{\rm Auger}\simeq 0.8 E_{\rm TA}$) and the differing trends of composition indicators at higher energies.  Currently the most extensive dataset on composition indicators, such as the average and the RMS of the depth of shower maximum  ($X_{\rm max}$), has been published by the Auger collaboration and shows a departure from a composition consistent with lighter nuclei at $10^{18}$ eV to a trend towards heavier nuclei above $10^{19}$ eV \cite{Abraham:2010yv,AugerIcrc11} (see Fig.~\ref{fig:Xmax}). TA reports shower behaviors consistent with protons \cite{TAicrc11}. The discrepancies in composition reports and the difference in absolute energy scale make it difficult to constrain proposed models for the origin of UHECRs. Fortunately, a cross-experiment effort to understand these discrepancies is currently on-going.

Here we show that the fast spinning pulsar birth model described in \cite{Blasi00,Fang12} can explain the observed spectrum (both the Auger and the TA spectra) and the composition trend observed by Auger  \cite{Abraham:2010yv,AugerIcrc11}. To fit these two observables we allow the freedom to vary the percentage of different elements that are accelerated in the pulsar wind 
divided into 3 groups: Hydrogen, Carbon group (CNO), and Iron. 
Although the surface of the rotating neutron star is thought to be a natural source of Iron, 
X-ray spectra of pulsars have shown evidence for Helium \cite{Sanwal02}, and  Carbon, Oxygen, and Neon 
\cite{Heinke10,Mori:2003bd}. In addition, the material leftover from the progenitor star is likely to be in the Carbon group (mostly Oxygen, see e.g., \cite{Woosley95,Arnett96,Woosley02,Dessart12}).
As we show below, to fit the Auger spectrum a balanced ratio between Hydrogen and CNO, with a  less fraction of Iron suffices, while to fit the TA spectrum a higher percentage of Iron is needed since the spectrum extends to higher energies (see Fig.~\ref{fig:UHEspec}).  The composition selected by the Auger spectrum gives a good fit to the Auger average shower maximum ($\left<X_{\rm max}\right>$) and the fluctuations around the mean (RMS($X_{\rm max}$)), see Fig.~\ref{fig:Xmax}. 

The contribution of young pulsars to the Galactic cosmic ray flux is related to the extragalactic contribution through the distribution of birth parameters. Here we calculate this contribution for different UHECR scenarios. The end of the Galactic spectrum is highly dependent on Galactic diffusion parameters  and the history of the most recent pulsar births in the Galaxy (such that  the flux in the VHE region varies over long time scales). However, the differences in the UHECR data, such as the different absolute energy scales of  the Auger and TA observatories, give rise to significant changes in the predictions for the Galactic component. The Auger fit implies  Galactic pulsars as the main contributors to VHECRs from $10^{18}\,$eV, while with the TA fit, they start dominating from $10^{17.5}\,$eV (see Fig.~\ref{fig:whole_spec}).  

As an acceleration model that naturally produces heavier nuclei at ultrahigh energies, the fast pulsar birth model illustrates the tension in fitting the UHECR spectrum, composition, and anisotropy measurements over the highest two decades in energy (for further discussion on this tension see, e.g. \cite{Allard05, Allard:2008gj, Hooper:2006tn, Hooper:2009fd, Taylor:2011ta, Aloisio:2009sj, Aloisio:2012ba, Arisaka:2007iz}). The relevant energy range is bracketed between a transition from Galactic to extragalactic and the cosmological GZK propagation effect. The observational uncertainties are still too large to allow for a clear picture of the source requirements. Great progress will occur when the leading experiments at the highest energies, Auger and TA, reach a better agreement on the absolute energy scale and the composition as a function of energy. In addition, increase statistics in observations, e.g., from the JEM-EUSO space mission \cite{Adams:2012tt}, can reveal a clearer anisotropy picture, which will greatly improve our ability to zero in on the best model for the origin of ultrahigh energy cosmic rays.

\section{Newborn Pulsars as cosmic ray sources}

The acceleration of particles in pulsar environments has been suggested since their discovery \cite{Gunn69}. While nearby pulsars show direct evidence of accelerated electrons and positrons,  the acceleration of hadrons is still unclear. The suggestion of pulsars as cosmic ray accelerators of VHECRs has been discussed in \citep{Karakula74,Bednarek97,Bednarek02,GL02,BB04} while for UHECRs the main proposals are by \cite{Blasi00,Arons03}. In \cite{Blasi00} Iron nuclei stripped off the neutron star surface are accelerated to UHEs by the fastest spinning young neutron stars with typical pulsar magnetic fields (between $10^{12}$ and $10^{13}$ G). They derived the maximum energy and spectrum due to the spin down of young pulsars ($J \propto E^{-1}$) that we use below. Neutron stars with much larger surface magnetic fields, i.e., magnetars, have been proposed as sources of ultrahigh energy protons by \cite{Arons03}. Given their faster spin down rate, the acceleration to UHEs is at earlier stages of the pulsar evolution when gravitational radiation is significant and a disruption of the supernova envelope is needed to allow the escape of accelerated particles \cite{Arons03,K11}. 

In \cite{Fang12} we discussed the acceleration and escape of UHE cosmic rays from newly born pulsars. Here we follow the same notation 
describing a pulsar with magnetic dipole moment $\mu=\mu_{30.5}\,10^{30.5}\,\rm cgs$ and rotational speed $\Omega=\Omega_4\,10^4\,\rm s^{-1}$,
that accelerate particles of charge $Z=26\,Z_{26}$ up to the energy   \cite{Blasi00}: 
\begin{eqnarray}\label{eq:Eacc}
E(\mu,\Omega, Z) \simeq 9\times10^{20}\,Z_{26} \, \eta_3  \,\Omega_4^2  \,\mu_{30.5}\; \textrm{eV}
\end{eqnarray}
where $0\le\eta\le 1$ is the pulsar acceleration efficiency, 
that is, the fraction of the full potential drop a charged particle experienced during its way out of the pulsar wind. We take $\eta_3\equiv{{\eta}/ {0.3}}$ in this work.

As the pulsar spins down, the energy of the particles produced in the wind decreases. A particle of energy $E$ is accelerated in the wind at time:
\begin{eqnarray}\label{eq:tspin_simple}
t_{\rm spin}(E) \simeq 3\times 10^{7}\, \left(\frac{9\times10^{20}\,\textrm{eV}}{E}\right) \frac{Z_{26}\eta_3I_{45}}{\mu_{30.5}}  \,{\rm s} \, ,
\end{eqnarray}
neglecting energy losses via gravitational waves (see \cite{Arons03,K11}). 

The cosmic ray spectrum injected over time by a pulsar with inertia $I=I_{45}\,10^{45} \,\rm g\,cm^2$ can then be calculated, assuming the Goldreich-Julian charge density, ${n}_{\rm GJ}= B\, \Omega /4 \pi Z ec$, \citep{Goldreich69} is entirely tapped in the outflow \cite{Blasi00}. The total charge density is the sum of the densities for each chemical group  $n_{\rm GJ} = \sum_Z  {\rm f}_Z  \, n_{\rm GJ}(Z)$, where ${\rm f}_Z$ is the fraction of the Goldreich-Julian charge density in particles with charge $Z$ injected into the pulsar wind and ${\rm f}_Z$ satisfies $\sum_Z {\rm f}_Z=1$. The injected spectrum of a given species $Z$ is then given by:
\begin{equation}\label{eq:spectrum_blasi}
\frac{{\rm d} N_{\rm inj}}{{\rm d} E }\,(\mu,\Omega, Z) = 5\times10^{23} \,{\rm f}_Z\, I_{45} (Z_{26} \, \mu_{30.5} \, E_{20})^{-1}\, \textrm{eV}^{-1}.
\end{equation}
In the pulsar model, the more energetic particles are accelerated at earlier times, when the surrounding supernova envelope is denser making the escape more difficult. Assuming a supernova expands with constant speed $v_{\rm ej}$ and explosion energy $E_{\rm ej}=E_{\rm ej,52}\,10^{52}\,\rm erg$, the mean density of the supernova ejecta with mass $M_{\rm ej, 10}=M_{\rm ej}\,10\,M_\odot$ is:
\begin{equation}\label{eq:rhoSN}
\rho_{\rm SN}(t)=\frac{M_{\rm ej}}{(4/3)\pi v_{\rm ej}^3t^3} \simeq 2\times 10^{-16} M_{\rm ej,10}^{5/2}E_{\rm ej,52}^{-3/2} t_{\rm yr}^{-3}~{\rm g\,cm}^{-3}\,\ ,
\end{equation}

Numerical results of the escaped spectrum can be found in Figure 3 of \cite{Fang12} for Hydrogen and Iron. Here we also include the results of similar calculations performed for the escape of injected Helium and CNO. The traversal of the supernova ejecta exponentially cuts the injection spectrum at: 
\begin{eqnarray}\label{eq:Ecut_estimate}
E_{{\rm cut},Z} & \simeq  & 2.3\times10^{19}\,Z_{1}\eta_3I_{45}\mu_{30.5}^{-1}M_{\rm ej,10}^{-1}E_{\rm ej,52}^{1/2}\left(\frac{\sigma_p}{\sigma}\right)^{1/2}\,{\rm eV} \label{eq:Ecut_proton} \\
&\simeq & 3.6\times 10^{20}Z_{26}\eta_3I_{45}\mu_{30.5}^{-1}M_{\rm ej,10}^{-1}E_{\rm ej,52}^{1/2}\left(\frac{\sigma_{\rm Fe}}{\sigma}\right)^{1/2}{\rm eV \, ,} \label{eq:Ecut_iron}
\end{eqnarray} 
where  $\sigma$ is the cross section of the hadronic interaction between cosmic rays and ejecta particles. In the relevant interaction energies, for proton-proton interactions it is approximately $\sigma_p\approx130\,$mb while it reaches  $\sigma_{\rm Fe}\approx1.25\,$b for Iron-proton interactions (note that we used cross sections with full energy and composition dependence in the simulations). 
For $E\le E_{{\rm cut},Z}$, the spectrum is not affected by the interactions. Above $E_{{\rm cut},Z}$, the escaped spectrum can be approximated by:
\begin{equation}
\frac{{\rm d} N_{\rm esc}}{{\rm d} E }\,(\mu,\Omega, Z) = \frac{{\rm d} N_{\rm inj}}{{\rm d} E }\,(\mu,\Omega, Z) \,e^{1-E/E_{\rm cut}}
\end{equation}

The chemical composition of the ejecta after steady and explosive burning of supernova II-P, Ib, and Ic has been studied by a number of authors,  e.g., \cite{Woosley95,Arnett96,Woosley02,Dessart12}.  Core-collapse supernova progenitors have large abundances of Oxygen produced by Helium burning during the life of the massive star, and part of this Oxygen is burnt during the explosion, which produces Si, So, Mg, Ca and Ni. 
Note that in most cases, Hydrogen is the dominant component in the envelope to consider for the escape of cosmic rays, as studied in \cite{Fang12}. The effect of escape from SN envelopes which are not proton dominated is small once the summing over all pulsars is considered as below.

The source of UHECRs in our model are the rare, extremely fast spinning, young pulsars. The majority of pulsars were born spinning slower and therefore contribute to the flux of lower energy cosmic rays. According to   \cite{Faucher06}, the distribution of pulsar birth spin periods, $f(P=2\pi/\Omega)$,  is normal, centered at 300 ms, with standard deviation of 150 ms.   We assume that proto-pulsars spinning initially below the minimum spin period of a neutron star $P_{\rm min}\simeq 0.6\,\rm ms$ \cite{Haensel99}  evolve to a stable figuration with $P=P_{\rm min}$. Hence we assume $f(P_{\rm min})=\sum_Pf(P<P_{\rm min})$.
The initial magnetic field follows a log-normal distribution $f(\mu)$ with $\langle \log (B/{\rm G}) \rangle\sim 12.65$ and $\sigma_{\log B}\sim 0.55$ \cite{Faucher06}.

The total cosmic ray spectrum contributed by the entire pulsar population is then
 \begin{equation}\label{eqn:sum_flux}
\frac{{\rm d} N}{{\rm d} E }\,(Z)=\int \frac{{\rm d} N_{\rm esc}}{{\rm d} E }\,(\mu,\Omega, Z) \,f(\mu) \,f(\Omega)\,d\mu\,d\Omega
\end{equation}
Cosmic rays injected by Galactic and extragalactic (EG) pulsars  travel through the Interstellar Medium (ISM) and the Intergalactic Medium (IGM) on their way to Earth. The corresponding propagation effects also affect the observed spectrum and composition and are discussed next in Section~\ref{sec:EG_propa} and Section~\ref{sec:G_propa}.

\section{Propagation from Extragalactic sources} \label{sec:EG_propa}

On average, the pulsar birth rate in the Galaxy is $\nu_{\rm s} \simeq 1/60\,{\rm yr}$ \citep{Lorimer08}. For a galaxy density  of $ n_{\rm gal}\simeq 0.02\,\rm Mpc^{-3}$ and an energy loss time for cosmic rays with energy
above $10^{19}\,\rm eV$ of $c\,T_{\rm loss}\sim 792 \,\rm Mpc$  \citep{Kotera11}, the average flux of cosmic rays from EG sources can be estimated to be: 

\begin{eqnarray}
\frac{{\rm d} N_{\rm EG}}{{\rm d} E \,{\rm d}t\, {\rm d}A\, {\rm d}\Omega}&=&\frac{{\rm d} N_{\rm }}{{\rm d} E }\,\frac{1}{4\pi}\,{c\,T_{\rm loss}}\,n_{\rm gal}\,\nu_{\rm s}\,f_{\rm s}\\
&=& \frac{{\rm d} N_{\rm }}{{\rm d} E }\,f_{\rm s}\,7 \times10^{-55}\,\rm eV^{-1}\,m^{-2}\,s^{-1}\,sr^{-1}
\end{eqnarray}

where $f_{\rm s}$ is an overall factor used to fit the model prediction to the measured UHECR flux. In the pulsar model, it can be interpreted as the  fraction of total flux of pulsar births required  to account for the observed flux of UHECRs out of the total pulsar birth rate. $f_{\rm s} < 1$ can be due to efficiency factors such as variations in the core-collapse geometry, poorer injection efficiency, or a lower hadronic density in the pulsar wind than the Goldreich-Julian density.

In order to estimate the observed spectrum of UHECRs, after their propagation through the extragalactic medium, we rescaled the simulation output of \cite{KAO10} from $10^{16}$ eV up to the maximum acceleration energy, $E_{\rm max}$ by the injected spectrum of pulsars.  The effect of the well-known Greisen-Zatsein-Kuzmin interactions of UHECRs with the cosmic microwave background \cite{G66,ZK66} (as well as other cosmic photon backgrounds) is included in these calculations, however, in the pulsar model, the GZK effect is secondary to the effect of $E_{\rm max}$ which is set by the fastest spin rate that neutron stars can reach $\Omega_{\rm max}\simeq 10^{4.2}\,\rm s^{-1}$ \cite{Haensel99}.

The UHECR spectrum and composition ratios for a given pulsar, after escape through the surrounding supernova ejecta, were calculated in \cite{Fang12}. At the highest energies, the spectrum is mostly determined by the superposition of exponential cutoffs up to $E_{\rm max}$, coming from the effect of escaping through the supernova ejecta. 
In order to account for the whole range of pulsar spins and magnetic fields, we ran $19\times19$ simulations of sets of $(P,\,\log{\mu})$, and integrated the obtained spectra over the overall pulsar population following Eq.~\ref{eqn:sum_flux}.

\begin{figure}[p]
\centering
\epsfig{file=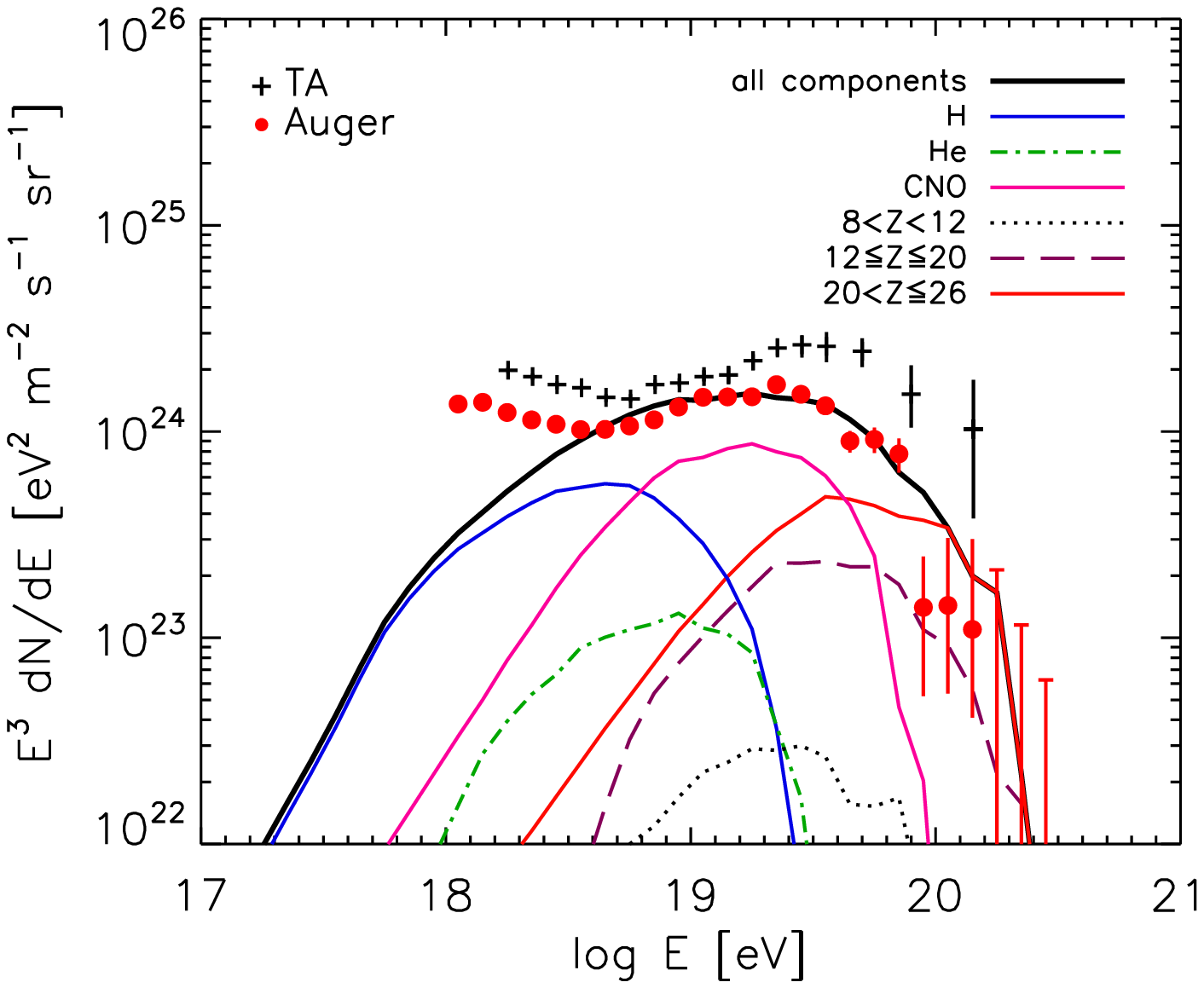,width=0.5\textwidth,clip=}
\epsfig{file=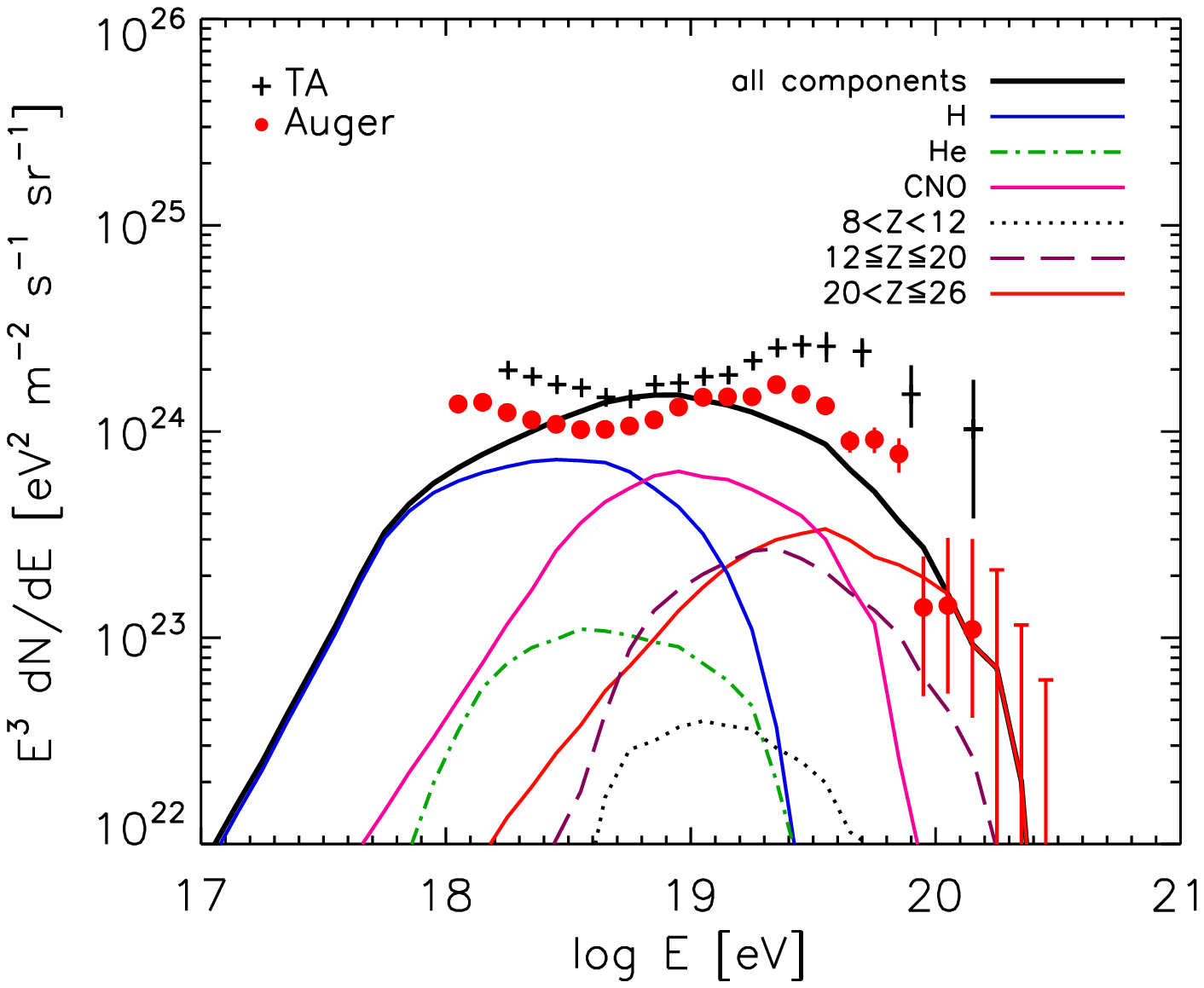,width=0.5\textwidth,clip=}
\epsfig{file=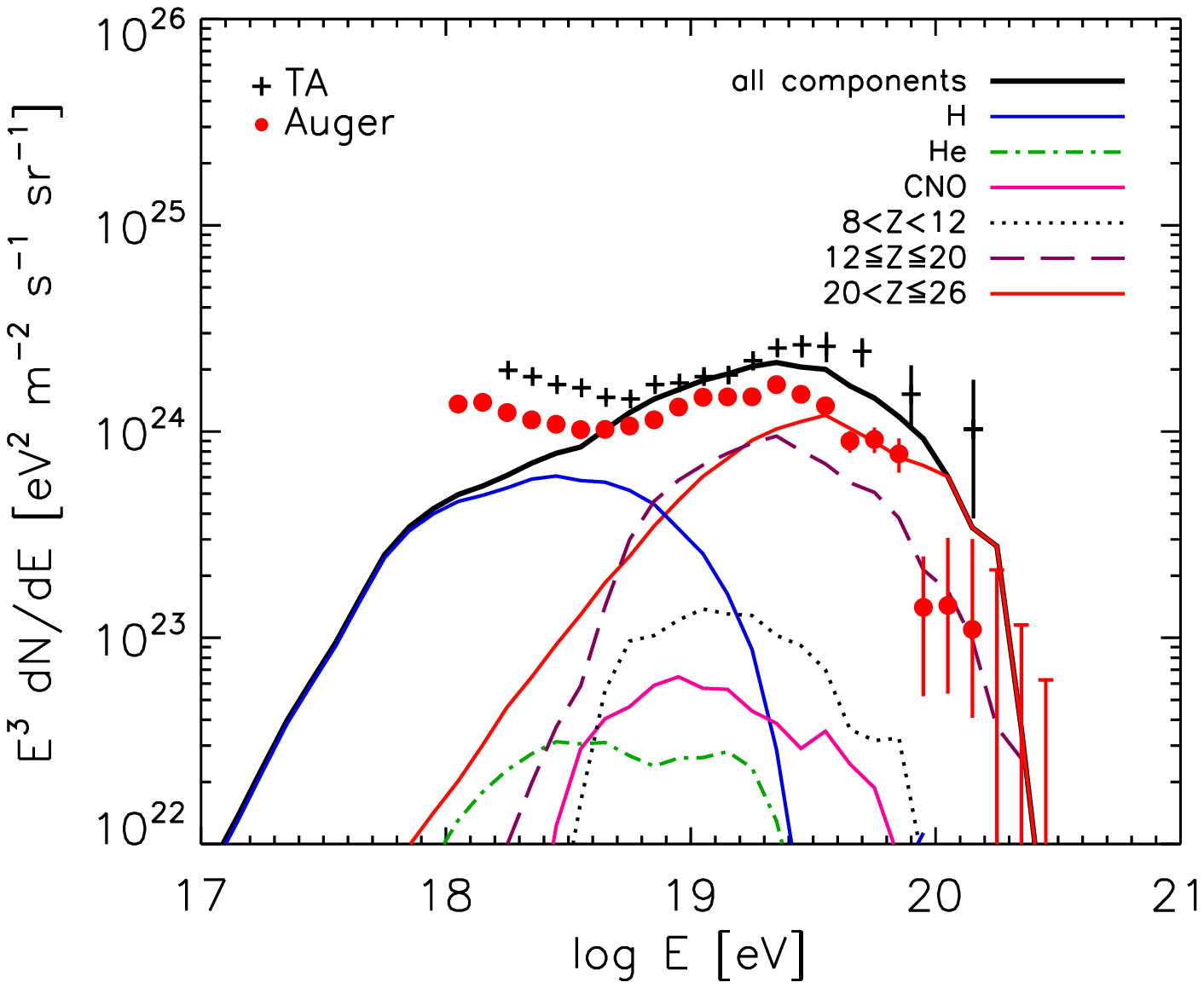,width=0.5\textwidth}
\caption{\label{fig:EG_spec} Propagated energy spectrum of UHECRs from newly born pulsar population with $\log{\mu}$ and $P$ normally distributed, and wind acceleration efficiency $\eta=0.3$.   Simulation results were normalized at $10^{19}\,\rm eV$ with $f_{\rm s}\simeq 0.05$ for the Auger and Auger-uniform cases, and $f_{\rm s}\simeq 0.08$ for the TA case (due to difference in energy scale). The spectrum of each group of propagated nuclei are shown as in the legend box. Top (Auger-uniform case): a mixed composition of $50\%$ Proton (${\rm f_{H}} = 0.5$), $30\%$ CNO   (${\rm f_{CNO}} = 0.3$) and $20\%$ Fe  (${\rm f_{Fe}} = 0.2$) was injected to fit the Auger spectrum \cite{Abreu:2011pj}. The source emissivity is assumed to be constant over time. Middle (Auger case): $65\%$ Proton, $20\%$ CNO and $15\%$ Fe nuclei was injected also to fit the Auger spectrum \cite{Abreu:2011pj}, but the source emissivity is assumed to be follow the star formation rate computed by \cite{Yuksel08}.   Bottom (TA case): $50\%$ Proton, $0\%$ CNO and $50\%$ Fe nuclei was injected to fit the TA spectrum \cite{Tsunesada:2011mp}. The source emissivity is assumed to be follow the star formation rate computed by \cite{Yuksel08}. 
\label{fig:UHEspec}}
\end{figure}

Figure~\ref{fig:EG_spec} shows the propagated energy spectrum of cosmic rays from EG pulsars. In the middle and bottom panels, the pulsar emissivity is assumed to follow the star formation rate computed by \cite{Yuksel08}. Note that a star formation history following the model of \cite{HB06} only changes slightly the results. For comparison, we also present a case where the source emissivity remains constant over time in the top panel, which we call Auger-uniform. (For a discussion on the influence of the source evolution model on the UHECR spectrum, see~\cite{KAO10}.)
In the Auger-uniform case, the sources were assumed to be uniformly distributed in space, normally distributed in $\log \mu$ and $P$ as in Equation~\ref{eqn:sum_flux} and with pulsar wind acceleration efficiency $\eta=0.3$. 

The overall normalization factor $f_{\rm s}\sim 0.05$ for the Auger-uniform (top panel) and Auger cases (middle panel),  and $f_{\rm s}\sim 0.08$ for the TA case (bottom panel), to take into account the difference in energy scale between the two experiments.
In the Auger-uniform case the injection composition is $50\%$ H (${\rm f_{H}} = 0.5$), $30\%$ CNO  (${\rm f_{CNO}} = 0.3$) and $20\%$ Fe  (${\rm f_{Fe}} = 0.2$). In the Auger case
${\rm f_H} = 0.65$, ${\rm f_{CNO}} = 0.20$, and ${\rm f_{Fe}} = 0.15$.
 These composition ratios were chosen to fit the Auger spectrum \cite{Abreu:2011pj} and composition \cite{Abraham:2010yv,AugerIcrc11}. In the TA case, an injection of ${\rm f_H} = 0.5$ and ${\rm f_{Fe}} = 0.5$ provides a better fit to the Telescope Array (TA) spectrum \cite{Tsunesada:2011mp}.

In the Auger case, our model over-produces the total flux between $10^{18.5}-10^{19}\,\rm eV$.
In this study we do not include the possible effect of a magnetic horizon as in \cite{AB05,L05,KL08a,Globus08} (see Section~\ref{sec:discussion}). This effect would harden the extragalactic spectrum around the ankle region, and thus enable a better spectral fit with the star formation rate scenario. It is also possible  that the extragalactic pulsar population that produces UHECRs has a biased evolution (see \cite{Fang12}) that can be weaker compared to the classical star formation rate. One would then obtain a better fit to the spectrum, as in the Auger-uniform case, but, as discussed below, to the expense of a good fit to the composition around the ankle region. 

\begin{figure}[p]
\centering
\epsfig{file=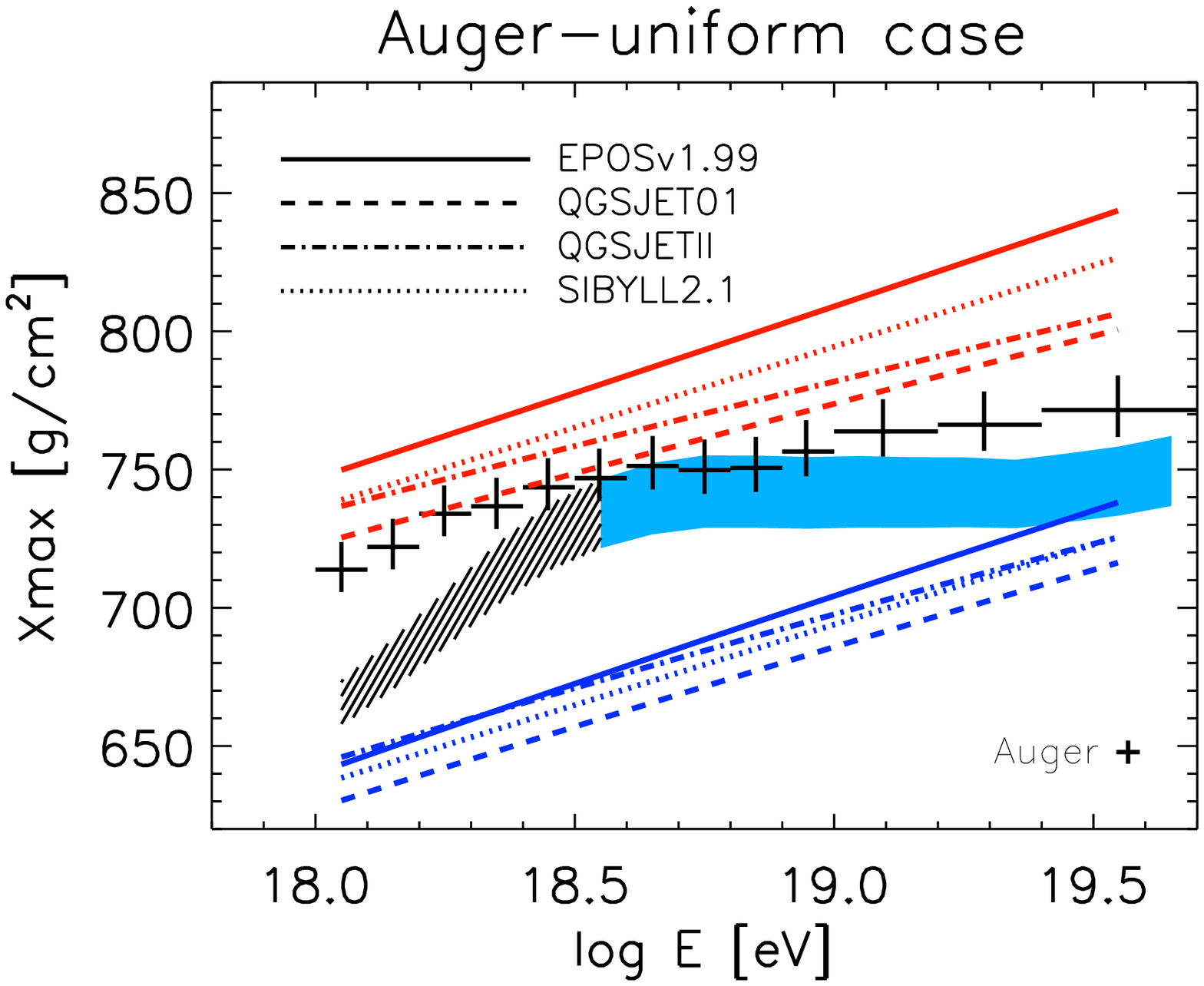,width=0.49\linewidth,clip=} 
\epsfig{file=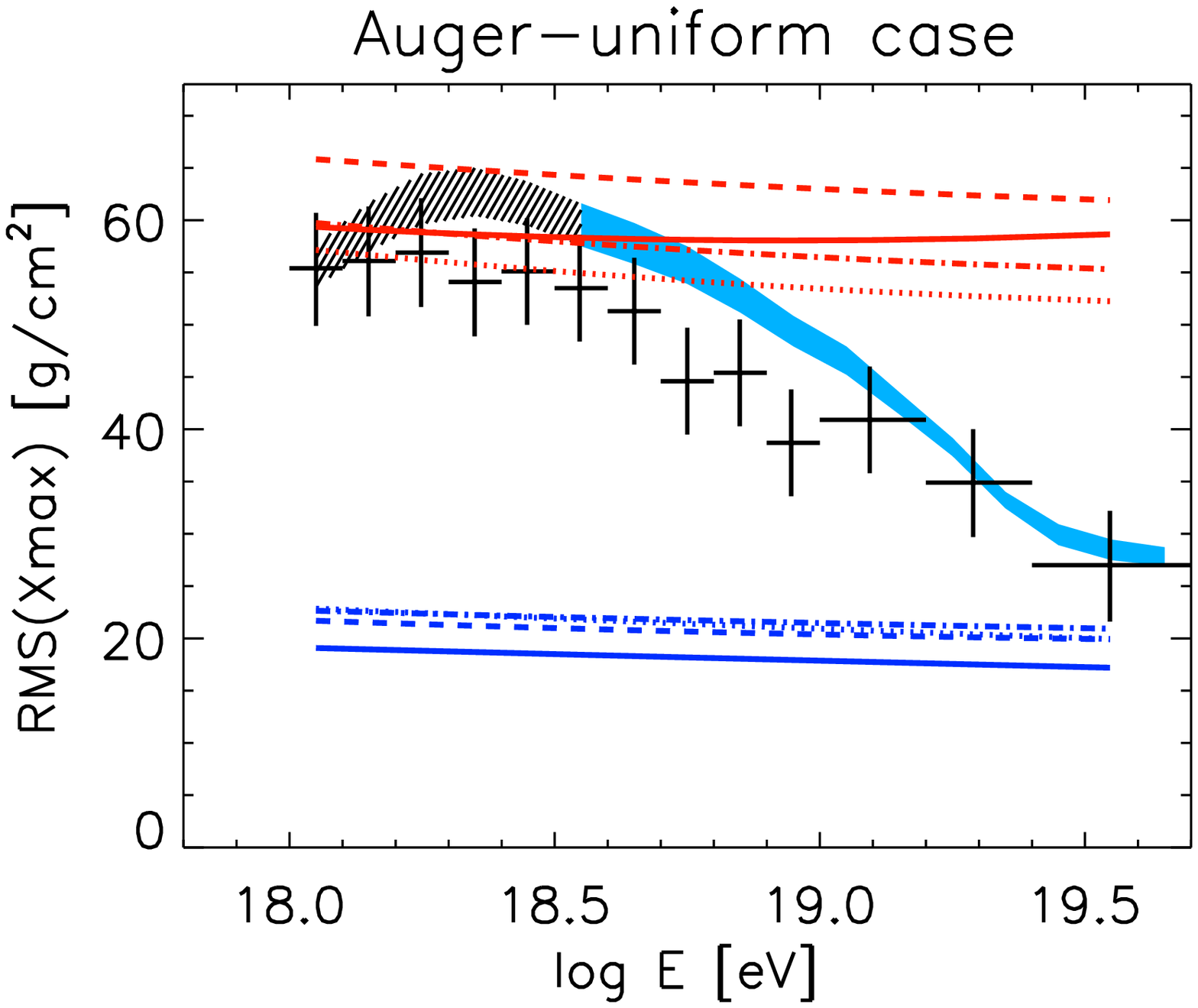,width=0.49\linewidth,clip=} 

\epsfig{file=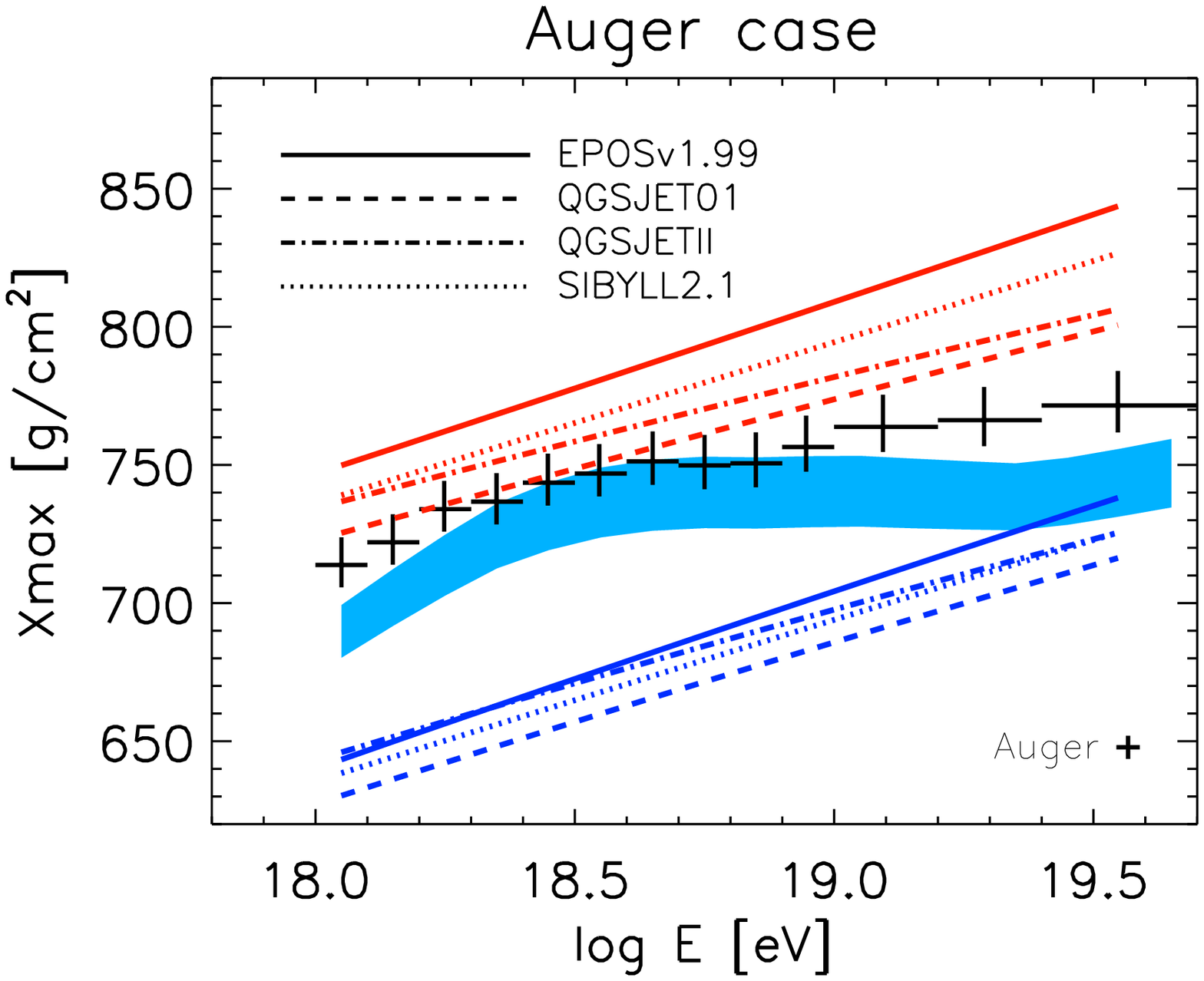,width=0.49\linewidth,clip=} 
\epsfig{file=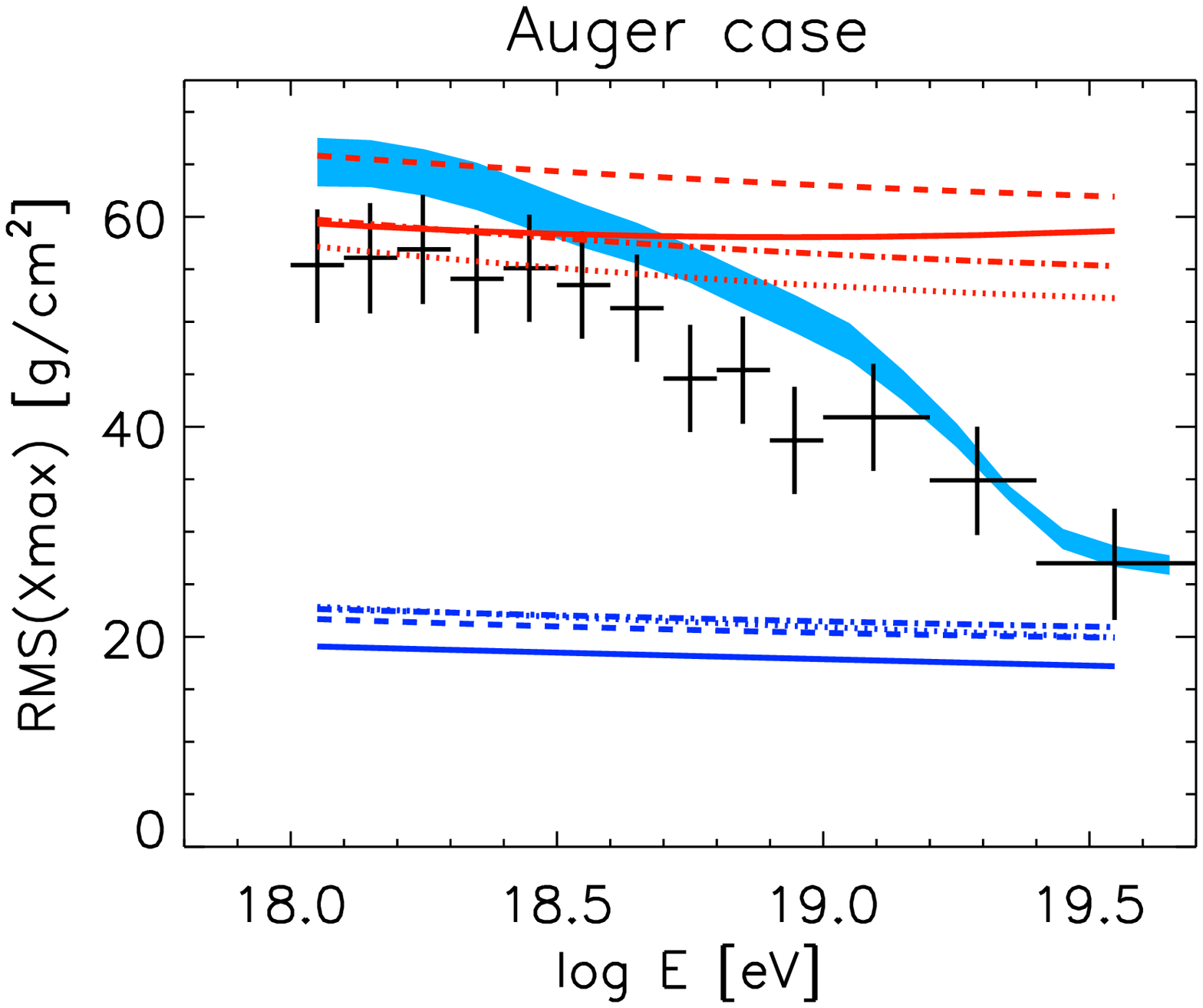,width=0.49\linewidth,clip=} 

\epsfig{file=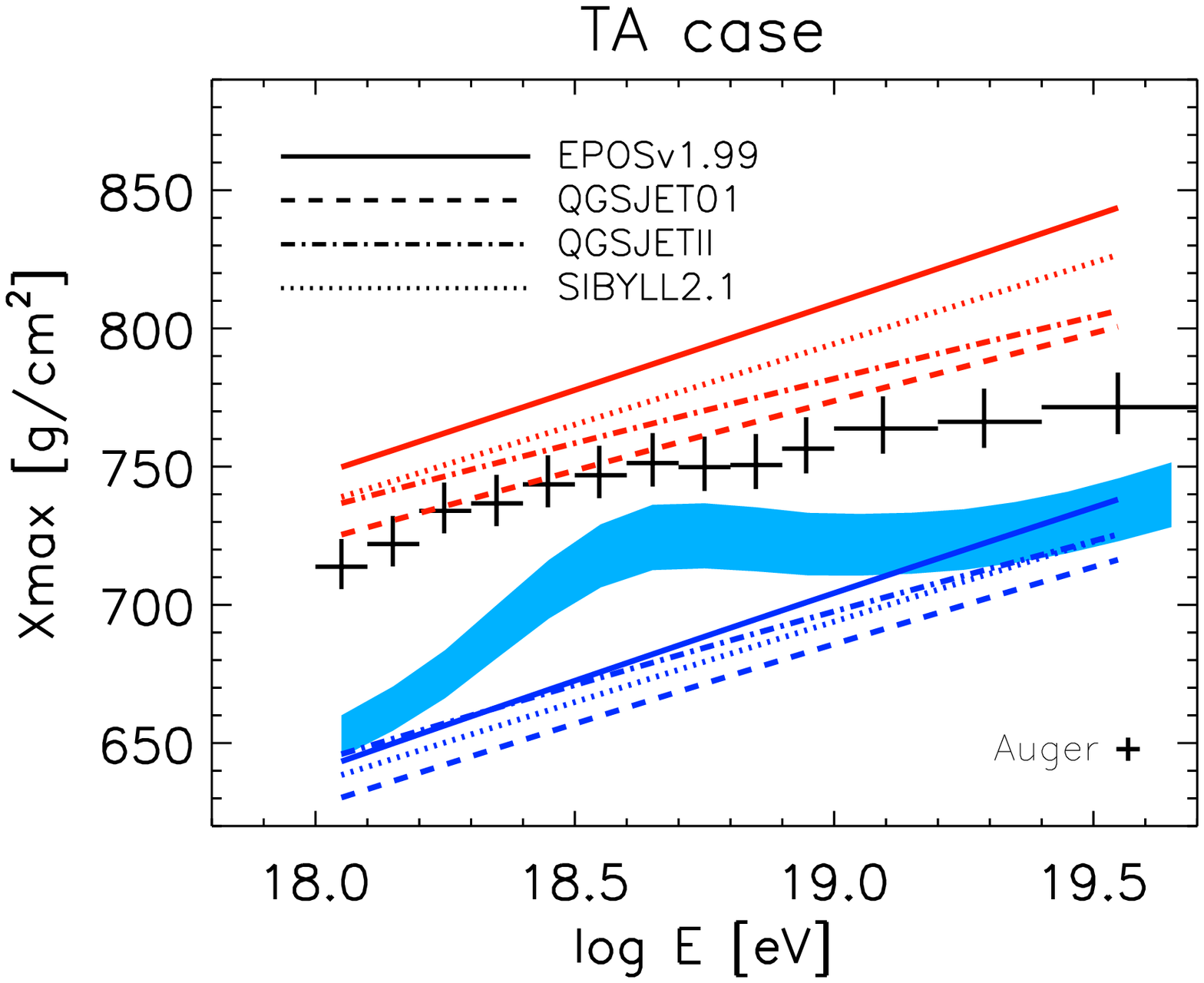,width=0.49\linewidth,clip=} 
\epsfig{file=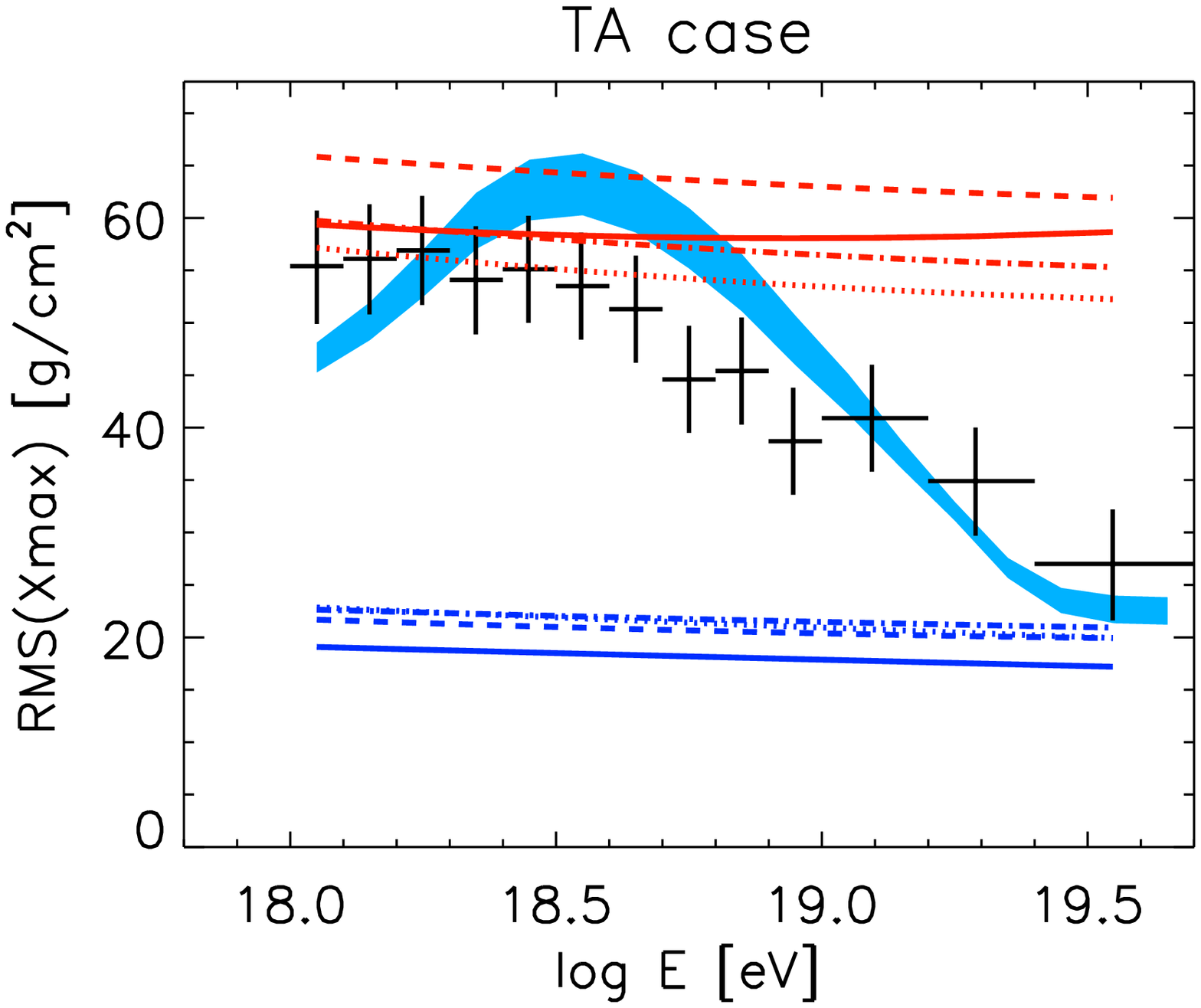,width=0.49\linewidth,clip=} 

\caption{\label{fig:EG_comp} $\langle X_{\rm max} \rangle$  (left column) and RMS($X_{\rm max}$) (right column) of the Auger  data~\cite{Abraham:2010yv,AugerIcrc11} (black crosses) and simulation results of the Auger-uniform (top), Auger (middle) and TA (bottom) cases as in Fig~\ref{fig:EG_spec}  (blue shaded region is for where pulsars contribute more than $80\%$ to the total flux, hashed region when they contribute less).
Flux from both Galactic and extragalactic pulsars are taken into account. Four interaction models, EPOSv1.99  \cite{Pierog:2006qv}, QGSJET01 \cite{Kalmykov:1997te}, QGSJETII \cite{Ostapchenko:2006wc} and SIBYLL2.1 \cite{Ahn:2009wx}  were used to estimate the range of $\langle X_{\rm max} \rangle$  and RMS($X_{\rm max}$)  as listed in the legend box. The red and dark blue lines correspond to $100\%$ P and $100\%$ Fe.
\label{fig:Xmax}}
\end{figure}

The mean atmospheric depth $\langle X_{\rm max} \rangle$ and its fluctuations, RMS($\langle X_{\rm max} \rangle$), of the Auger case compared with Auger measurements \cite{Abraham:2010yv} are presented in Fig.~\ref{fig:EG_comp}.  Four hadronic interaction models EPOSv1.99  \cite{Pierog:2006qv}, QGSJET01 \cite{Kalmykov:1997te}, QGSJETII \cite{Ostapchenko:2006wc} and SIBYLL2.1 \cite{Ahn:2009wx} were used in this calculation, which gives the range of the blue shaded region (also see \cite{Abreu:2013env} for fits to the Auger data with different hadronic interactions). Our estimates include the contribution of Galactic pulsars, as calculated in the following section. 

Our results fit the observations better with the EPOS model. (Note that we used the EPOS model in our simulation of the escape of accelerated particles from the supernova ejecta.) The Auger and Auger-uniform cases follow well the trend of a transition from light to heavy elements as measured by the Auger Observatory. In the Auger case, the composition is dominated by the contribution of extragalactic pulsars down to $E\sim 10^{18}\,$eV. In the Auger-uniform case however, pulsars become an underdominant source below $E\sim 10^{18.5}\,$eV, as the extragalactic contribution cuts off at the ankle, and the Galactic pulsar population provides only a fraction of the total flux below. As a result, the composition below the ankle appears heavier than measured, and another light component is needed (which raises some questions about the origin of this other component as discussed at the end of the Section~\ref{section:VHE}).
Under our chosen assumptions, the TA case does not fit the constant light composition measurements of TA and is a bit heavier than the Auger composition.

\begin{figure}[p]
\centering
\epsfig{file=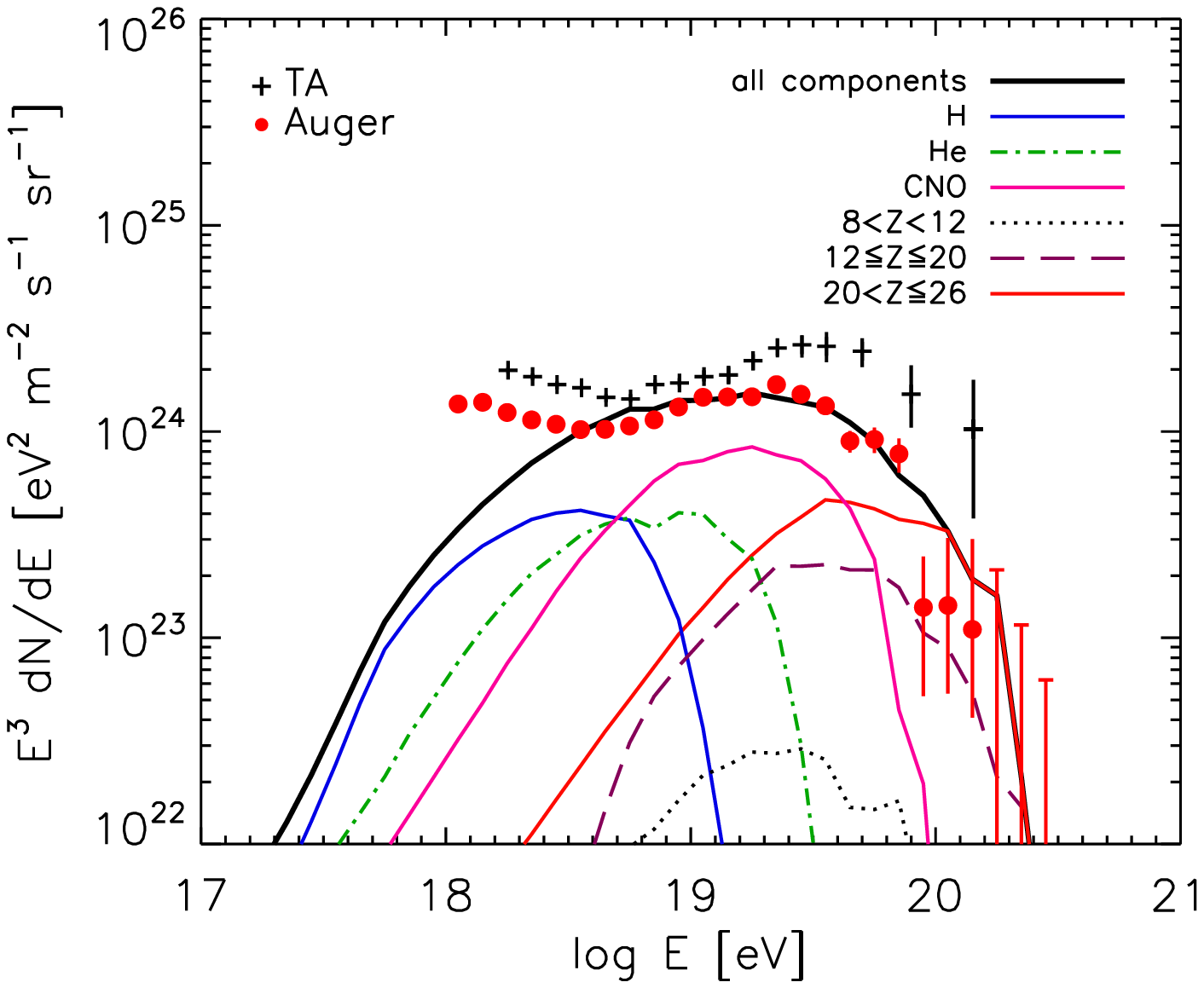,width=0.5\textwidth,clip=}
\epsfig{file=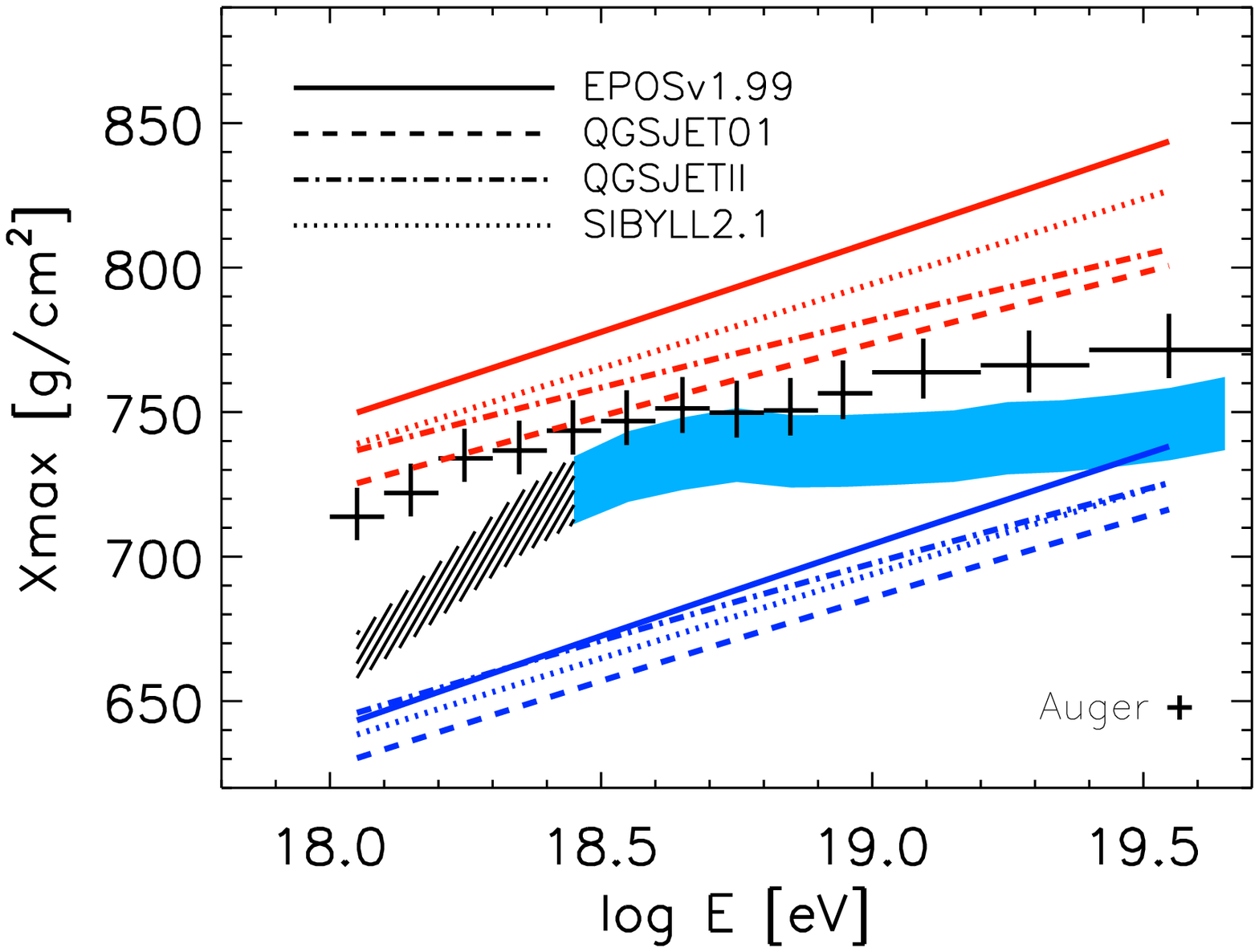,width=0.5\textwidth,clip=}
\epsfig{file=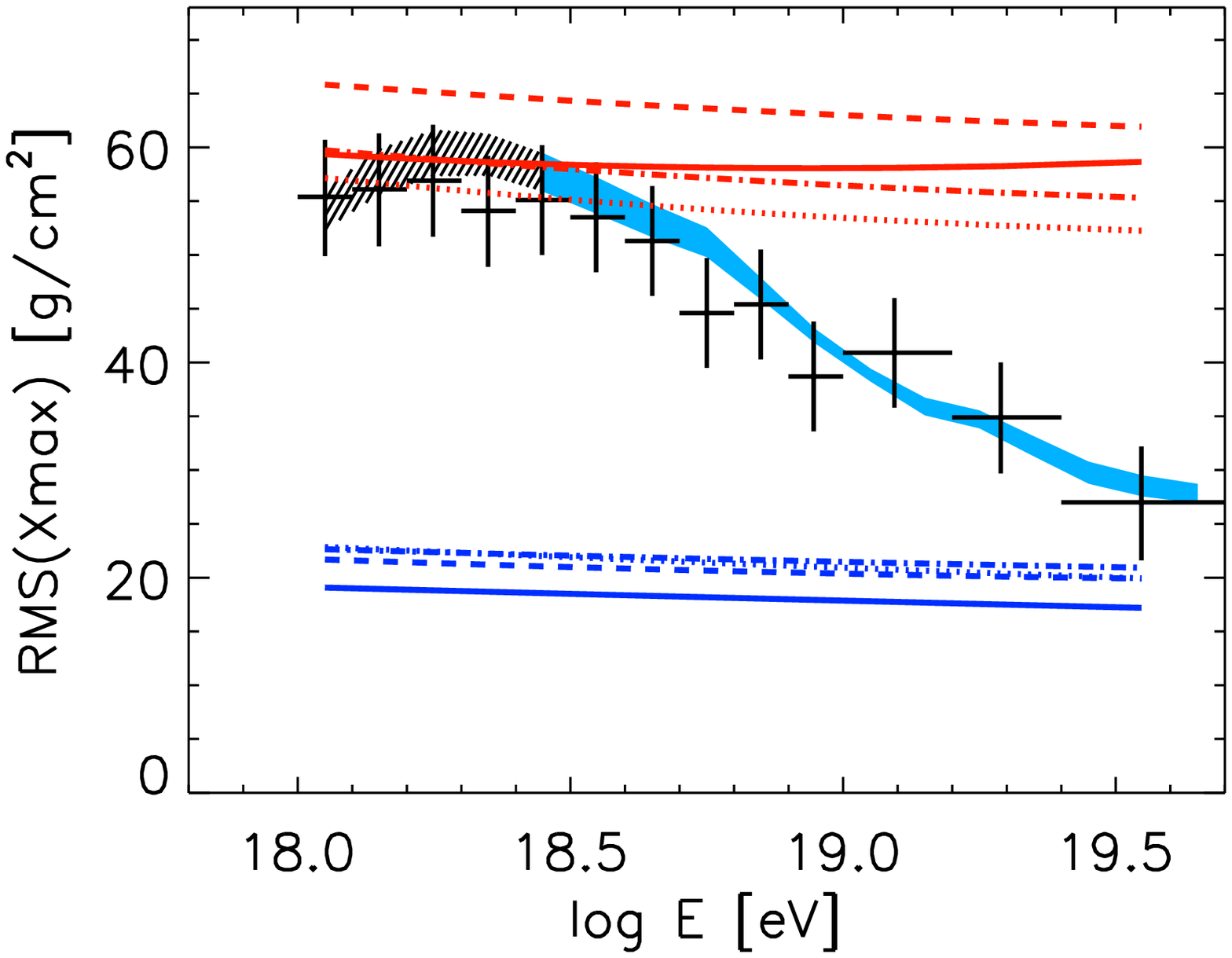,width=0.5\textwidth}
\caption{\label{fig:He} Propagated energy spectrum (top), $\langle X_{\rm max} \rangle$  (middle) and RMS($X_{\rm max}$) (bottom) of an alternative Auger-uniform case, with Proton interchanged to Helium at injection. A mixed composition of $50\%$ Helium (${\rm f_{He}} = 0.5$), $30\%$ CNO   (${\rm f_{CNO}} = 0.3$) and $20\%$ Fe  (${\rm f_{Fe}} = 0.2$) was injected to fit the Auger spectrum \cite{Abreu:2011pj}. The source emissivity is assumed to be constant over time. Newly born pulsar population are assumed to have $\log{\mu}$ and $P$ normally distributed, and wind acceleration efficiency $\eta=0.3$.   Simulation results were normalized at $10^{19}\,\rm eV$ with $f_{\rm s}\simeq 0.05$. Notice the propagated spectrum and composition remain almost unchanged compared to the Auger-uniform case with $50\%$ Proton, $30\%$ CNO   and $20\%$ Fe injection (top panels in Fig~\ref{fig:UHEspec} and Fig~\ref{fig:Xmax}).     }
\end{figure}

Because protons dominate the flux at ankle energies, and  Helium nuclei are mostly dissociated into protons during the propagation, the injected protons  can be mostly interchanged to Helium without affecting the spectrum significantly. For example in the Auger-uniform case, the composition and spectrum after propagation remains nearly unchanged for an injection of ${\rm f_H} = 0$, ${\rm f_{He}} = 0.5$, ${\rm f_{CNO}} = 0.3$, and ${\rm f_{Fe}} = 0.2$ (Figure~\ref{fig:He}). 

\section{Propagation from Galactic sources} \label{sec:G_propa}

For the propagation of cosmic rays accelerated by Galactic pulsars, we model the turbulent Galactic magnetic field as a cylindrical halo of radius $R_{\rm Gal}=15\,$kpc, of height above (or below) the Galactic plane typically $H\sim 2-8\,$kpc \cite{Mao12}, of coherence length $l_{\rm c}=10-100\,\rm pc$ and strength $B=3\,\mu$G (see \cite{Han08} and references therein). The Larmor radius of a particle reads
${r_{\rm L}}=13.8\,E_{18}\,Z_{26}^{-1}\,(B/3\,\mu\rm G)^{-1}\,{\rm pc}$.

The flux of cosmic rays accelerated by Galactic pulsars can then be calculated as
\begin{eqnarray}\label{eqn:gal_flux}
\frac{{\rm d} N_{\rm Gal}}{{\rm d} E \,{\rm d}t\, {\rm d}A\, {\rm d}\Omega}&=&\frac{{\rm d} N_{\rm esc}}{{\rm d} E }\,\frac{1}{4\pi}\,\frac{c\,\delta t_{\rm obs}}{V_{\rm Gal}}\,\nu_{\rm s}\,f_{\rm s}\,\left(1-e^{-\delta t_{\rm obs}/\delta t_{\rm birth}}\right) \, ,
\end{eqnarray}
where $f_{\rm s}$ is the same scale factor introduced for the EG component, $\nu_{\rm s}$ is the average birth rate of pulsars in our Galaxy, and $V_{\rm Gal}=2\pi\,HR_{\rm Gal}^2$ is the volume of the Milky Way. The average number of sources in the Galaxy that contributes to the observed spectrum at a given energy $E$ is $N_{\rm s}(E)=\delta t_{\rm obs}(E)/\delta t_{\rm birth}(E)$, where $\delta t_{\rm obs}(E)$ is the timescale over which a source can can contribute to the observable cosmic rays of energy $E$, and $\delta t_{\rm birth}(E)^{-1}$ is the birth rate of Galactic sources that can produce particles of energy $E$. 
Following a Poisson distribution, the probability that currently at least one source is contributing in the Galaxy  is $(1-e^{-\delta t_{\rm obs}/\delta t_{\rm birth}})$.

If the scattering length of the particle (distance over which its deflection angle becomes $\delta \theta \sim 1$) is shorter than the height of the Galaxy, the propagation will be mostly diffusive, and the source observation time is equivalent to the particle escape time from the Galaxy: $\delta t_{\rm obs}=\tau_{\rm esc}$. At energies above the knee ($\sim 10^{15}$ eV), which is the main concern of this paper, nuclei spallation is negligible \citep{Blasi12a} and the nuclei escape time can simply be estimated with the Leaky box model
\begin{equation}\label{eqn:esc}
\tau_{\rm esc}(E, Z, l_{\rm c})=\frac{H^2}{2D} \, , 
\end{equation}
where $D$ is the diffusion coefficient in the Galactic magnetic field, that can be estimated empirically as in Equation A2 of \cite{Kotera08a}
\begin{equation}\label{Eqn:diff}
D (E,Z, l_{\rm c})\sim D_0\,
   r_{\rm L}\,c\,\left[\frac{r_{\rm L}}{l_{\rm c}}+\alpha\,\left(\frac{r_{\rm L}}{l_{\rm c}}\right)^{-2/3}\right]. 
   \end{equation}
The coefficient $\alpha$ depends on the turbulence and structure of the Galactic magnetic field. The normalization $D_0$ is set at energies where particles are in the Kolmogorov regime (${r_{\rm L}}\ll {l_{\rm c}}$), using measurements of the boron to carbon ratio in our Galaxy. We follow the estimates of \cite{Blasi12a}:
\begin{eqnarray}
D(R)=1.33\times10^{28}\,D_0\,H_{\rm kpc}\,\left(\frac{R}{3\,\rm GV}\right)^{1/3}\,\rm cm^2s^{-1} \, ,
\end{eqnarray} 
where $H_{\rm kpc}=H/1\,{\rm kpc}$, and the particle rigidity $R\equiv E/Z$.
Notice that in the diffusive regime, the diffusion coefficient scales as $E^{-\beta}$, with $\beta=1/3$, a value which seems to be favored by observations, e.g., \cite{Blasi12a}. 
A larger value of $\beta$ would imply a faster escape out of the Galaxy of particles at very high energies. For example, $\beta=0.6$ would imply that all particles above $4\times10^{15}\,\rm GV$ travel rectilinearly, which is inconsistent with  anisotropy measurements \citep{Kifune86,Aglietta09,Amenomori05}.

For scattering lengths larger than $H$, the propagation is quasi-rectilinear. In principle, the observation time $\delta t_{\rm obs}$ is then equivalent to the dispersion of particle arrival times $\sigma_t = D_{\rm s}\,\delta\theta^2/(4c)$, where $D_{\rm s}$ is a typical source distance \cite{AH78,Harari02b,KL08b}. The transition from a totally diffusive regime to a quasi-rectilinear regime can be modeled using Eq.~\ref{Eqn:diff} with $\alpha=1$. For reasonable assumptions on the structure of the magnetic field, one can then assume $\delta t_{\rm obs}=\sigma_{\rm t}\sim \tau_{\rm esc}$ at high rigidities. 

\section{Very High Energy and Ultrahigh Energy cosmic rays from Pulsars}\label{section:VHE}

\begin{figure}[h]
\centering
\epsfig{file=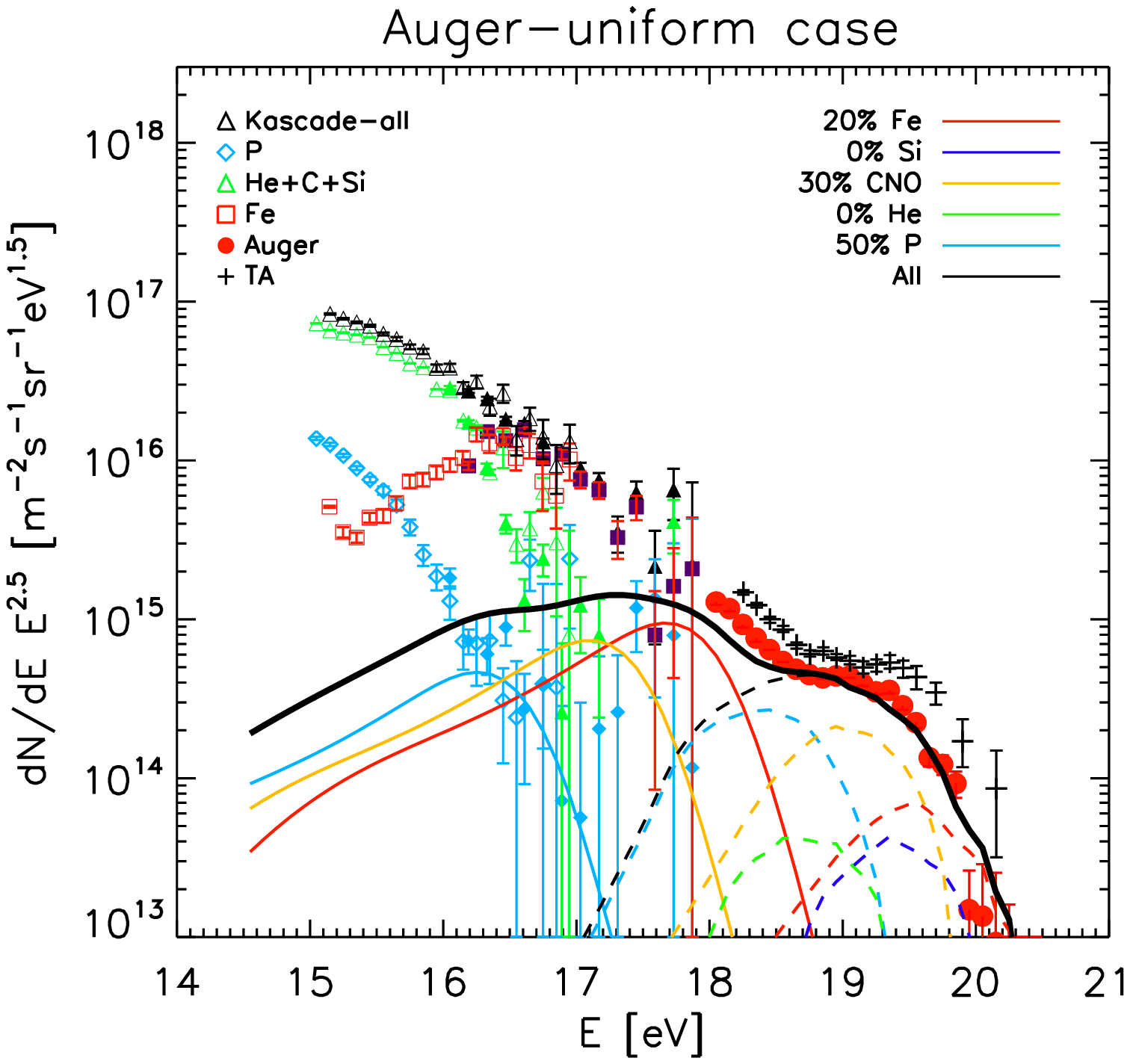,width=0.5\linewidth,clip=} 
\epsfig{file=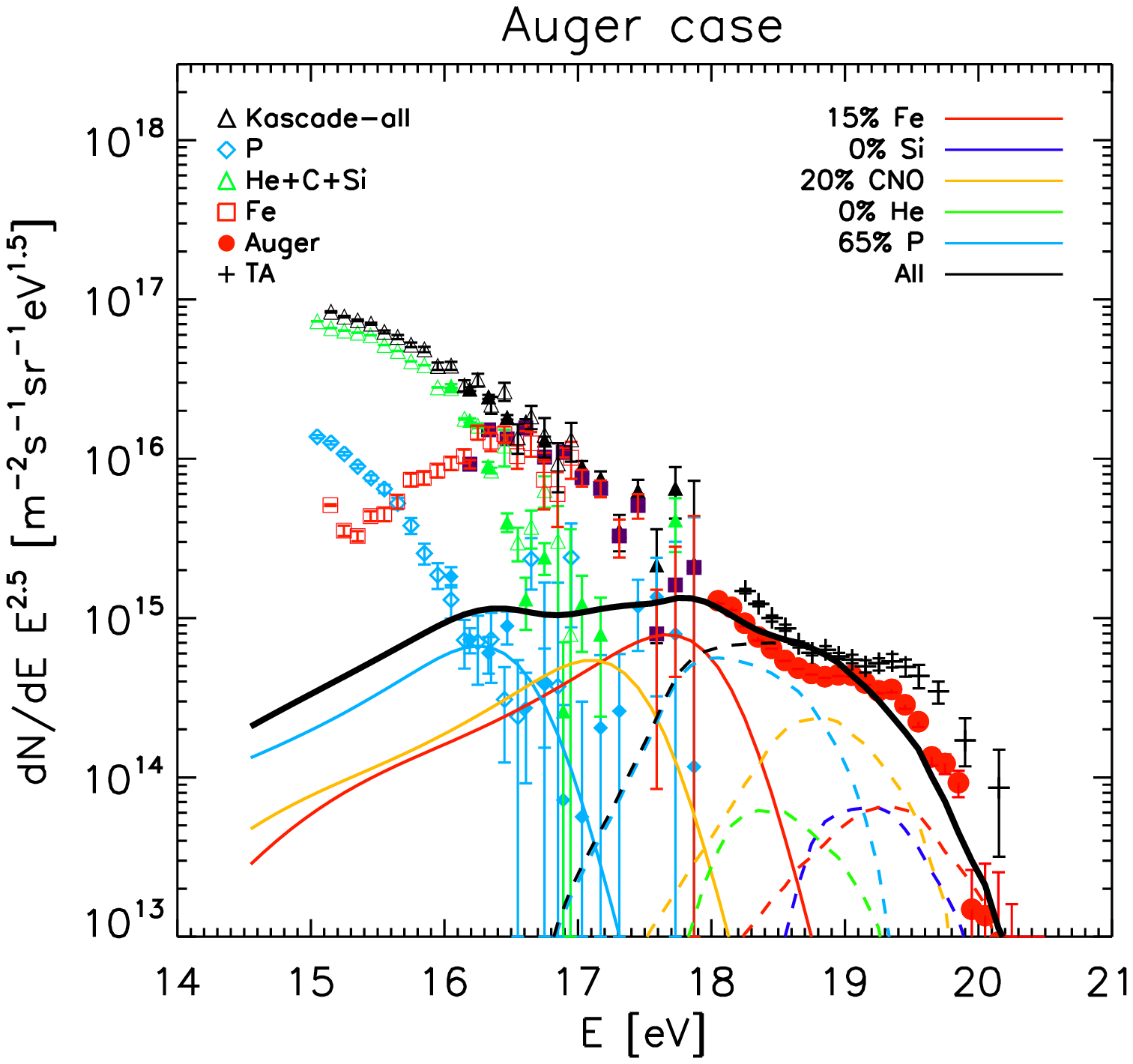,width=0.5\linewidth,clip=} 
\epsfig{file=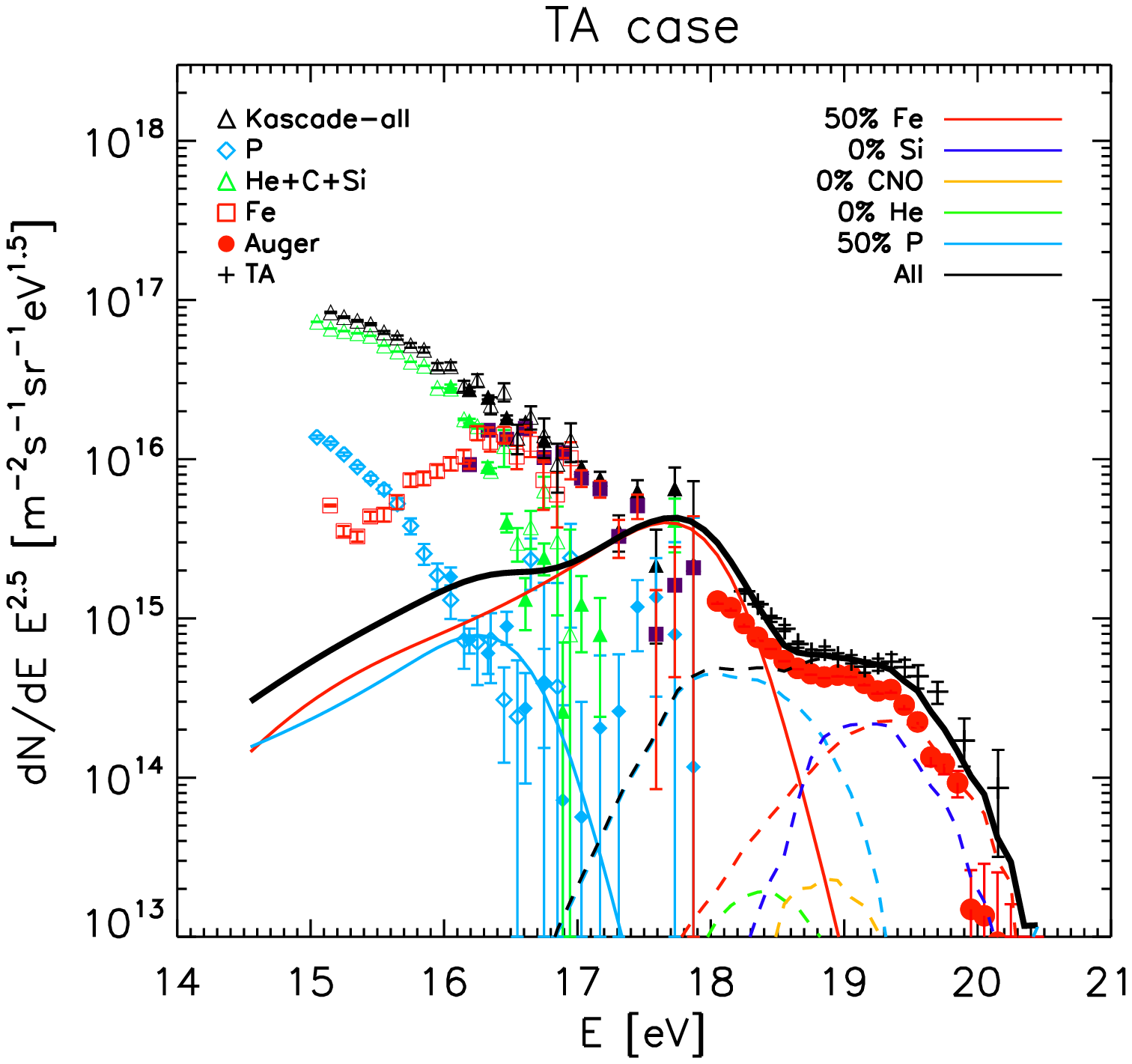,width=0.5\linewidth,clip=} 
\caption{\label{fig:whole_spec} Cosmic ray flux measurements by KASCADE-Grande \cite{Apel:2011mi},  Auger \cite{AugerIcrc11} and TA \cite{Jui:2011vm} compared with pulsar model predictions. The total spectrum in solid black sums up extragalactic (dash) and Galactic (solid) components.
The top, middle, and bottom panels correspond to the Auger-uniform, Auger, and TA cases respectively, as in Fig~\ref{fig:EG_spec}. Pulsar and propagation parameters: wind acceleration coefficient $\eta=0.3$, Galactic magnetic field coherence length $l_c=20\,\rm pc$, magnetic halo height $H=2\,\rm kpc$. 
\label{fig:whole_spec}}
\end{figure}

Assuming that cosmic rays injected in the ISM by Galactic pulsars have the same composition as those from  EG pulsars,  we can calculate the contribution of both Galactic and extragalactic pulsars. Figure \ref{fig:whole_spec} show the results for the Auger, Auger-uniform (top and middle panels) and the TA cases (bottom panel)  where we also show the total energy spectrum and the spectrum decomposed into three components (Hydrogen, intermediate, and Iron) as reported by {KASCADE} \cite{Antoni:2003gd}, {KASCADE-Grande} \cite{Apel:2011mi} and the Auger and TA spectra. The EG spectra (in dash lines) in Fig.~\ref{fig:whole_spec} are the same as those in Fig.~\ref{fig:EG_spec}. 

The Galactic spectrum of each element group (in solid lines) has three regimes. For example, if we consider the Iron branch in the  Auger-uniform case (top panel), it  
shows a totally diffusive behavior between  $E=10^{14.5}\,\rm eV$ and $10^{17.5}\,\rm eV$ (where $r_{\rm L}=l_{\rm c}$). In this range the Galactic propagation softens the intrinsic spectrum by $1/D(E)\sim E^{-0.3}\,  Z^{0.3}$.   The second regime lies roughly between $10^{17.5}\,\rm eV$ and $10^{19}\,\rm eV$, where $r_{\rm L}>l_{\rm c}$ and particles random walk with small deflections leaving the Galaxy a bit faster, so  $D(E)\sim E^{2}\,Z^{-2}$  (see Equation~\ref{Eqn:diff}). However the second regime is overwhelmed by the last component, which comes in above $10^{18}\,\rm eV$ when particles escape  faster than the rate they are born, resulting in an event probability of $(1-e^{-\delta t_{\rm obs}/\delta t_{\rm birth}})$ and a time dependent flux. The three branches with $Z=1, 7, 26$ for Hydrogen, CNO, and Iron have similar behaviors along the energy axis scaled by $Z$, i.e., same behavior for the same rigidity.  The relationship between the  amplitudes of the flux for each component combines the injection and propagation.  The injected spectrum is inversely proportional to the charge $dN_{\rm esc}/dE\propto Z^{-1}$. If each of the three elements are  injected with fractions $F(Z)$, then $dN/dE\propto Z^{-0.7}\,F(Z)$ in the totally diffusive regime and $dN/dE\propto Z^{1}\,F(Z)$ in the small deflection regime. 

\begin{figure}[h]
\centering
\epsfig{file=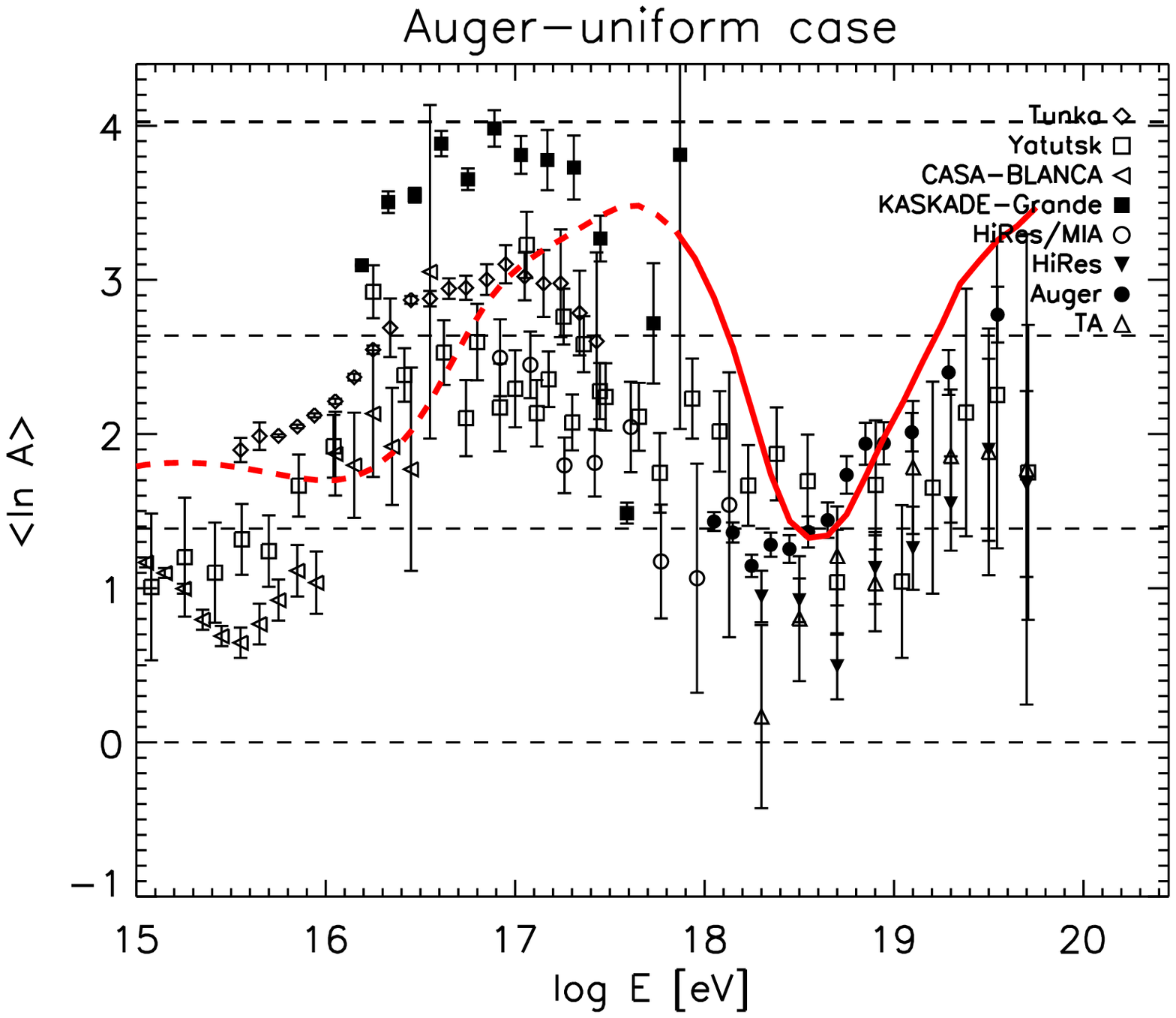,width=0.5\linewidth,clip=} 
\epsfig{file=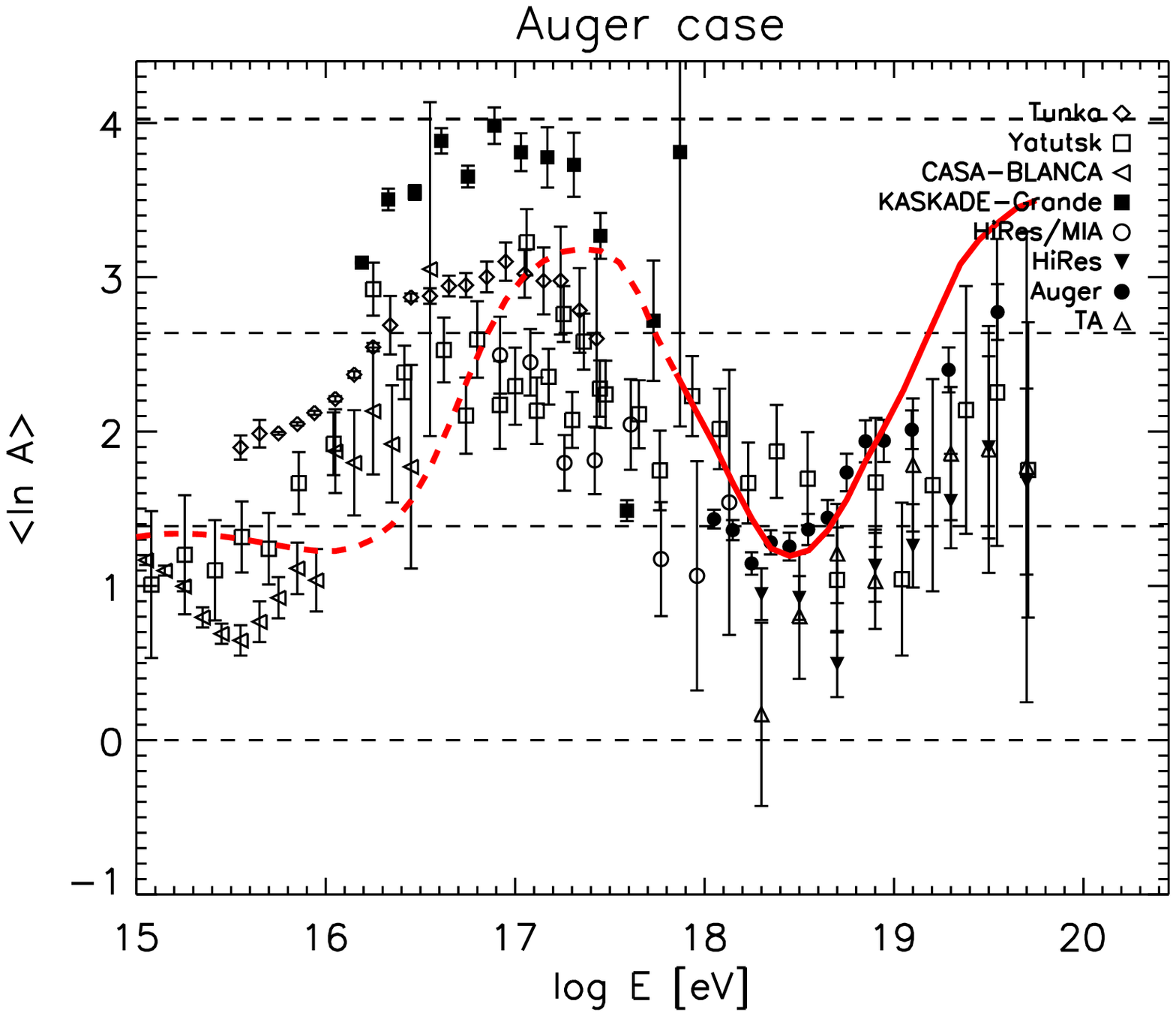,width=0.5\linewidth,clip=} 
\epsfig{file=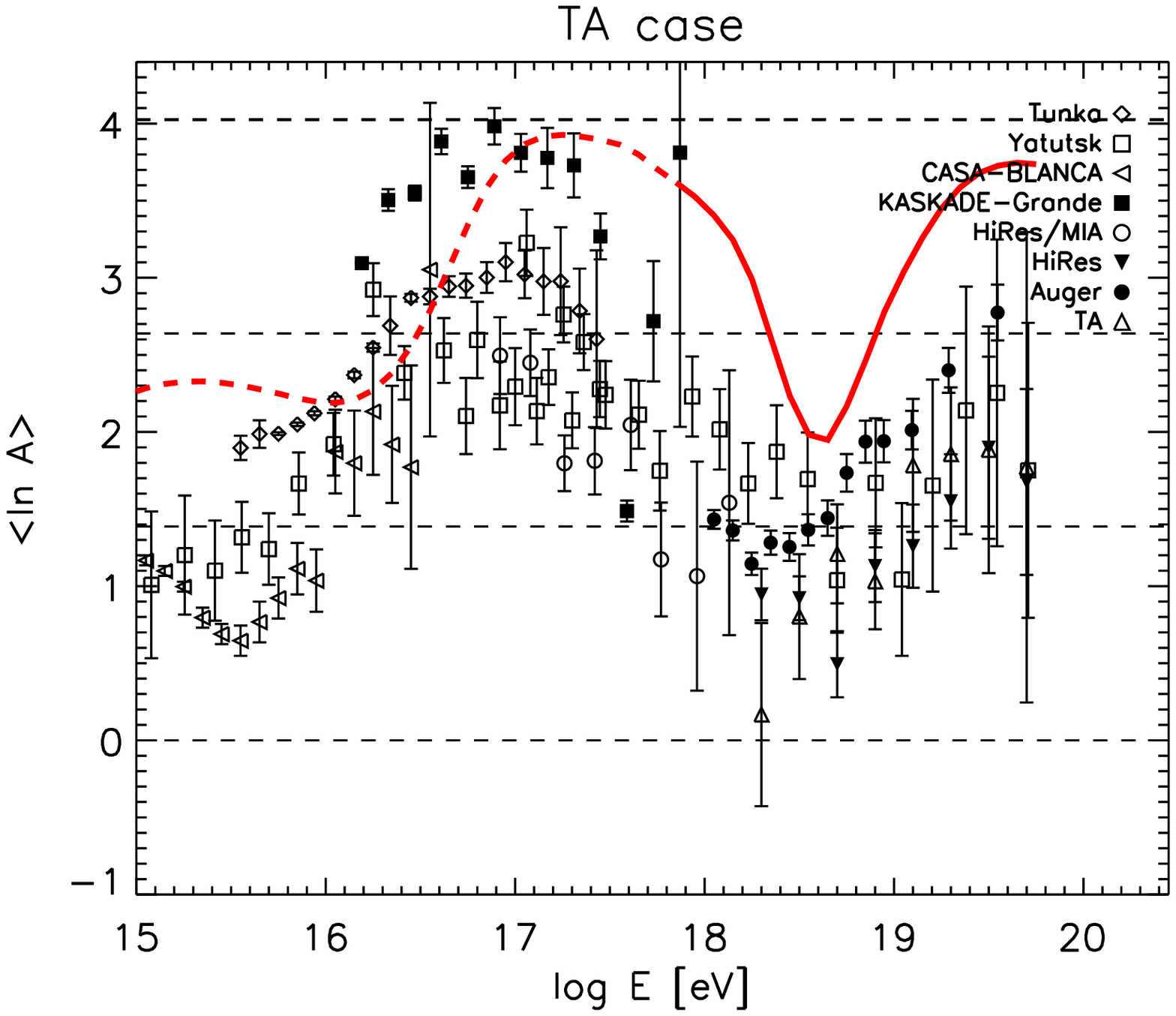,width=0.5\linewidth,clip=} 
\caption{\label{fig:lna} Average logarithmic mass of cosmic ray derived from $X_{\rm max}$ measurements from \cite {Kampert:2012mx} with data from Tunka \cite{Budnev:2009mc}, Yakutsk \cite{Knurenko:2010eu, Knurenko:2011zz}, CASA-BLANCA \cite{Fowler:2000si}, HiRes/MIA \cite{AbuZayyad:2000ay}, HiRes \cite{Abbasi:2009nf},  KASCADE-Grande \cite{Apel:2011mi}, Auger \cite{AugerIcrc11} and TA \cite{Jui:2011vm} for hadronic interaction model EPOS v1.99 \cite{Pierog:2006qv} compare with simulation predictions (red lines) as in Fig.~\ref{fig:whole_spec}. Dashed lines indicate the energy range where pulsars contribute less than $80\%$ to the total flux (see Fig.~\ref{fig:whole_spec}) and other sources also contribute.  
\label{fig:lnA}}
\end{figure}

Figure~\ref{fig:lna} contrasts the composition predictions of our models with the measurements of $\ln{A}$ from
Tunka (\cite{Budnev:2009mc}, filled square), Yakutsk (\cite{Knurenko:2010eu, Knurenko:2011zz}, open square), CASA-BLANCA (\cite{Fowler:2000si}, open left triangle), HiRes/MIA (\cite{AbuZayyad:2000ay}, open circle), HiRes (\cite{Abbasi:2009nf}, filled downward triangle), KASKADE-GRANDE(\cite{Apel:2011mi}, filled diamond), Auger(\cite{AugerIcrc11}, filled circle) and TA(\cite{Jui:2011vm}, open upward triangle) based on hadronic interaction model EPOSv1.99 \cite{Pierog:2006qv}. The composition trends of our model reproduces some of the Galactic-Extragalactic transition features at the ankle as shown by the Auger and the KASKADE measurements. The contribution below the ankle depends on the dominant component in this energy range {(defined here as more than $80\%$ flux contribution)} which may or may not be the pulsars (see e.g the Auger-uniform case). If the additional flux comes form acceleration in supernova remnants the composition is likely to be mostly Iron below the ankle (e.g., \cite{Ptuskin10}).

The expected anisotropy signal at the highest energies (above $E_{\rm GZK}\equiv 6\times 10^{19}\,$eV) was already discussed in \cite{Fang12}. Because extragalactic pulsars are difficult to detect beyond our Local Group, and because of their transient nature, no direct correlation should be found between the arrival directions of the most powerful events and active sources. The distribution of arrival directions in the sky should trace the large scale galaxy distribution with a possible bias \cite{KW08,KL08b}, if the deflections experienced by the particles are small. At energies well below $E_{\rm GZK}$, the arrival directions of particles emitted in extragalactic sources should be isotropic whatever their composition, as the deflections experienced by protons are already large enough to prevent any clustering.

Nevertheless, Galactic sources producing protons above energies $\sim10^{17}\,$eV could produce substantial anisotropies.
In the regime of diffusive propagation, the anisotropy signal in a given direction can be defined as $\delta_{\vec{x}}=(\nabla_{\vec{x}}n_{\rm CR}/n_{\rm CR})\,(3\,D/c)$, where $n_{\rm CR}$ is the cosmic ray number density in the Galaxy as measured at the position of the Earth. Reference~\cite{Blasi12b} showed that by assuming a homogeneous distribution of sources in the Galactic disc, the small-scale anisotropy signal can be simplified to (see also \citep{Ptuskin06, 2012arXiv1208.5338P})
\begin{equation}\label{eq:anisotropy}
\delta=\frac{3}{2^{3/2}\,\pi^{1/2}}\,\frac{D(E)}{Hc}\ .
\end{equation}
%Figure~\ref{fig:anisotropy} shows the predicted anisotropy signal originated from the Galactic pulsar population 
Assuming diffusion parameters $H=2\,\rm kpc$ and $l_{\rm c}=20\,\rm pc$, a strong anisotropy  above rigidity $R\sim 4\times10^{17}\,\rm V$ would be expected from the non-diffusive regime when particles travel semi-rectilinearly. Measurements around $10^{18}\,$eV indicate that cosmic rays at these energies are mostly light (see, e.g., KASCADE-Grande data), and are distributed isotropically in the sky \cite{Kifune86,Aglietta09,Amenomori05}. The combination of these measurements require that particles around $10^{18}\,$eV (below the ankle) be protons accelerated in extragalactic sources. Galactic sources injecting protons at these energies would indeed induce strong anisotropies. This is a difficulty for most ``ankle"-transition models, for which the extragalactic component becomes quickly underdominant below $E_{\rm ankle}\equiv 10^{18.5}\,$eV. This is the case in particular for the Auger-uniform model, in which the transition at $E_{\rm ankle}$ implies that another light extragalactic component is needed to bridge the gap between a Galactic (e.g., SNR) component and the extragalactic pulsar contribution.

In the Auger and TA cases however, for energies roughly below $10^{16}\,\rm eV$, Galactic pulsars have a negligible contribution to the anisotropy signal, as other Galactic sources (e.g. SNR) dominate; for energies above $10^{18}\,\rm eV$, extragalactic sources dominate, which should produce a globally isotropic signal. For the energy range in between, Fig.~\ref{fig:whole_spec} shows that the cosmic ray flux is mostly composed of CNO and Iron nuclei, that produce less anisotropy as they have smaller rigidity than protons at the same energy. In addition, \cite{Giacinti11, Blasi12b} show that effects of stochasticity in the spatial and  temporal distribution of sources and  turbulent magnetic field could cause significant fluctuation to the intrinsic anisotropy. In general our results are consistent with the anisotropy measurements \citep{Kifune86,Aglietta09,Amenomori05}. 

This study reveals the tension between the spectrum, the composition, the anisotropy signal of cosmic rays above $10^{16}\,$eV and the need to have more precise measurements of these observables. The Auger-uniform model, though it can fit successfully the observed spectrum and the composition, requires a second extragalactic component of light elements around $10^{18}\,$eV. The Auger case fits successfully the composition and alleviates the anisotropy issue at $10^{18}\,$eV, but one has to invoke a strong magnetic horizon effect to harden the spectrum at $E_{\rm ankle}$. Finally, the TA case is satisfactory for spectrum and anisotropy, but does not seem to account well for the observed composition.

%\begin{figure}[h]
%\centering
%\epsfig{file=anisotropy.ps,width=0.7\linewidth,clip=} 
%\caption{{Small-scale anisotropy signal of cosmic rays observed at Earth (Eq.~\ref{eq:anisotropy}), calculated with the diffusion coefficient chosen in this paper and Galactic magnetic field parameters: $H=2\,\rm kpc$ and $lc=20\,\rm pc$, as a function of rigidity. Note that the effects of stochasticity of the spatial and temporal distribution of sources may cause significant fluctuations from this theoretical expectation.} }
%\label{fig:anisotropy} 
%\end{figure}

\section{Discussion}\label{sec:discussion}

The results that we provided here for the Galactic component, depend on our choice of parameters. 
Our fits to the observables of low energy cosmic rays were performed by adjusting key parameters of the Galactic magnetic fields: the coherence length of its turbulent component, $l_{\rm c}$, and the height of its halo, $H$. Our knowledge on these parameters remain poor (\cite{Beck08} for a review), and leaves room to other shapes and normalizations for the Galactic component calculated here. For example, for $H$ higher (lower) than 2~kpc, the field would confine $H/{2\, \rm kpc}$ times more (less) nuclei, as $dN_{\rm Gal}/dE\propto H^2/D(E)\propto H$.  The Galactic flux would hence move up (down) by $H/{2\,\rm kpc}$. A larger (smaller) coherence length of the field would enable particles with higher (lower) energy to be confined, and would shift the Galactic component to higher (lower) energies. Note that this effect is non linear, according to Equation~\ref{Eqn:diff}.  Further uncertainties in the modeling of particle diffusion, as discussed extensively in Ref.~\cite{Blasi12a}, could also noticeably modify the shape of the Galactic spectrum. 
We also assumed a uniform density of sources in our Galaxy to estimate the average properties, but at the very high energies, cosmic ray observables depend on the particular history of pulsar births in the Galaxy.

On the other hand, our results on the UHE component due to the extragalactic pulsar population can be regarded as fairly robust. For UHECRs indeed, the shape of the propagated spectrum depends mostly on the injected spectrum and composition, the source emissivity evolution (see e.g., \cite{KAO10, KO11}), and possibly on the presence of nearby sources at the highest energy end. 

The effects of the extragalactic magnetic fields are not taken into account in the propagation of UHECRs. At energies around the ankle, the presence of a mean field of order $1\,$nG could make the trajectories of particles so diffusive that they would not be able to propagate over the Hubble distance. This ``magnetic horizon" effect could harden the spectrum of $\sim E^{0.1-0.3}$ around the ankle region (e.g., \cite{L05, AB05, KL08a, Globus08}). This effect would flatten the bump in the ankle region that appears in Fig.~\ref{fig:UHEspec} and Fig.~\ref{fig:whole_spec}, and lead to a satisfactory fit of our model to the data for a broader range of Galactic magnetic field parameters. 

Note that the normalization factor of $f_{\rm s}\approx0.05$ that we require for a good fit, is a physically reasonable value. This factor accounts for pulsar wind injection efficiencies and for the fraction of the total pulsar population producing cosmic rays. A value of $f_{\rm s}\sim 0.05$ can mean that only 5\% of the pulsars are required to have the right configuration to produce cosmic rays. Assuming the gaussian distribution of pulsar parameters proposed by \cite{Faucher06}, less than 0.3\% of these 5\% are  pulsars born with periods less than $6\,\rm ms$, that can accelerate particles to above $10^{19}\,$eV (these estimates can be modulated by the injection efficiencies). 

The present results also depend on the distribution function of pulsar parameters. We normalized here the overall Galactic and extragalactic spectra based on the data at the highest energies, i.e., with the fastest-spinning pulsars corresponding to the tail of the distribution. Slight variations in the bulk of the population could hence impact the shape of the spectrum, and thus the composition.
For example, if $f(P)$ cuts at $P_{\rm min}$ instead of having those spin faster piled up to $P_{\rm min}$, to meet the observed flux level at UHE, the Galactic component in the Auger case would over-produce VHECRs measured by Kascade.

Recently, Ref.~\cite{Klein12} discussed that muons produced by hadronic and photopion interactions during the acceleration process could be accelerated significantly before their decay, producing an enhanced neutrino flux at high energies. The authors consequently claim that most UHECR candidate sources involving unipolar induction should be ruled out, as they would overproduce neutrinos compared to the IceCube limits, via this mechanism. However, UHE ions injected at low latitude in rotation powered pulsars or magnetars barely meet background hadrons while they `surf-ride' in the force-free magnetosphere \cite{Contopoulos02, Arons03}. Furthermore, we pointed out in Ref.~\cite{Fang12}, that the radiation fields in the pulsar wind are too low to impact the acceleration of UHECRs (see also \cite{Protheroe98}, \cite{Bednarek02}). Note also that Ref.~\cite{Klein12} assumes an injection in $E^{-2.3}$ by linear acceleration in their calculations, while  our induction model  implies a harder injection in $E^{-1}$ (Eq.~\ref{eq:spectrum_blasi}). This harder injection should result in a lower neutrino flux at the highest energies, than estimated by Ref.~\cite{Klein12}. It  thus appears that the muon acceleration mechanism should be negligible in our framework.

Our scenario, as for any viable scenario, assumes that {\it no} recent events in our Galaxy that can reach $E\gtrsim 10^{18.5}\,$eV, in order not to overshoot the observed spectrum, and not to produce a striking anisotropy pattern in the sky. If a nearby pulsar were born with high enough spin to reach $\gtrsim 10^{18.5}\,$eV, then we should see a surge in the cosmic ray flux after a delay of thousands of years (for a similar treatment of GRBs see \cite{2011ApJ...738L..21E, 2011ApJ...742..114P}).

\section{Conclusion}

We showed how fast spinning pulsars can explain the observed spectrum of UHECRs (both Auger and TA) and the composition trend described by the Auger collaboration. To fit these two observables a mixed composition of Hydrogen, CNO, and Iron needs to escape the young supernova remnants accelerated via the fast spinning pulsar winds. To fit the Auger spectrum a balanced ratio between Hydrogen and CNO, with a minor presence of Iron suffices, while to fit the TA spectrum a higher percentage of Iron is needed. Determining the absolute energy scale is thus an important goal for current observatories, as it would help select between possible explanations for the origin of UHECRs. 

The composition mixture chosen to fit the Auger spectrum, gives a very good fit to the average shower maximum ($\left<X_{\rm max}\right>$) and the fluctuations around the mean (RMS($X_{\rm max}$)) observed by Auger. This is a unique aspect of the pulsar model as most models of astrophysical accelerators of UHECRs do not explain these shower maximum data. This challenge has led to the notion that the change in $\left<X_{\rm max}\right>$ and RMS($X_{\rm max}$) is due to new physics in hadronic interactions at these energies, which are well above those reached by the Large Hadron Collider.

Another aspect of this model worth highlighting is the connection between parameters needed to fit the extragalactic component and the presence of a Galactic component from Galactic pulsar births in the very high energy range (between $10^{16}$ and $10^{18}$ eV). 
%If the composition in this transition region is better determined and the propagation of VHECRs in the Galaxy better constrained, then the required injection can be determined by VHECRs with a clear prediction for the energy scale and composition of UHECRs. Conversely, the energy scale and composition of UHECRs, can determine the relevance of the pulsar birth acceleration for Galactic cosmic rays. 
In the estimates presented here,  the Auger-uniform fit implies an under dominant contribution to the flux of VHECRs, while the Auger and TA fits suggest that Galactic pulsars could be the main contributors to VHECRs. %At VHEs the cosmic ray flux from pulsar births is time dependent and can vary by orders of magnitude if a nearby pulsar is born spinning fast.   

\acknowledgments
We thank E. Amato, P. Blasi, J. Cronin, L. Dessart, F. Ionita, S. Phinney, and P. Privitera for very fruitful discussions. This work was supported by the NSF grant PHY-1068696 at  the University of Chicago, and the Kavli Institute for Cosmological Physics through grant NSF PHY-1125897 and an endowment from the Kavli Foundation.
KK and AVO acknowledge financial support from PNHE.

\bibliography{FKO13}

\vfill\eject
%\end\bye
\end{document}